\documentclass[a4paper,13pt]{article}
\usepackage[utf8]{inputenc}
\usepackage{graphicx}
\usepackage{color}
\usepackage{array}
\newcolumntype{C}[1]{>{\centering\let\newline\\\arraybackslash\hspace{0pt}}m{#1}}
\usepackage{floatrow}
\usepackage[usenames,dvipsnames]{xcolor}
\usepackage{tikz}
\usetikzlibrary{shapes,arrows}
\usetikzlibrary{decorations.markings}
\usetikzlibrary{decorations.pathreplacing}
\usepackage{amsmath}
\usepackage{amsfonts}
\usepackage{amssymb}
\usepackage{cite}
\usepackage{float}
\usepackage{changepage}
\usepackage{nicefrac}
\usepackage{multirow}
\usepackage{booktabs}
\newcommand*\rot{\rotatebox{90}} 

\usepackage{geometry}
\title{Spatio-temporal Models of Lymphangiogenesis in Wound Healing}

\author{Arianna Bianchi, Kevin J. Painter, Jonathan A. Sherratt}

\date{2016}

\begin{document}

\maketitle

\noindent
\textbf{ABSTRACT:} \emph{Several studies suggest that one possible cause of impaired wound healing is failed or insufficient lymphangiogenesis, that is the formation of new lymphatic capillaries. Although many mathematical models have been developed to describe the formation of blood capillaries (angiogenesis), very few have been proposed for the regeneration of the lymphatic network.
Lymphangiogenesis is a markedly different process from angiogenesis, occurring at different times and in response to different chemical stimuli. Two main hypotheses have been proposed: 1) lymphatic capillaries sprout from existing interrupted ones at the edge of the wound in analogy to the blood angiogenesis case; 2) lymphatic endothelial cells first pool in the wound region following the lymph flow and then, once sufficiently populated, start to form a network.
Here we present two PDE models describing lymphangiogenesis according to these two different hypotheses. Further, we include the effect of advection due to interstitial flow and lymph flow coming from open capillaries. The variables represent different cell densities and growth factor concentrations, and where possible the parameters are estimated from biological data.
The models are then solved numerically and the results are compared with the available biological literature.}

\section{Introduction}

The lymphatic system first came to the anatomists' attention with Hippocrates' mention of lymph nodes in his 5\textsuperscript{th} century BC work \emph{On Joints} \cite{withington1984}.
Later, the Roman physician Rufus of Ephesus identified the axillary, inguinal and mesenteric nodes and the thymus in the 1\textsuperscript{st}-2\textsuperscript{nd} century AD \cite{may1968}. The earliest reference to lymphatic vessels is attributed to the anatomist Herophilus, who lived in Alexandria in the 3\textsuperscript{rd} century BC; he described the lymphatics as ``absorptive veins'' \cite{dodson1924,staden1989}. This rudimentary knowledge of the lymphatic system was lost during the Middle Ages, until Gabriele Falloppio re-discovered lymphatic capillaries in the mid-16\textsuperscript{th} century \cite{castiglioni1947}. Since then, there has been a steady but slow increase in our awareness of the ``second'' circulatory system of our body (see \cite{ambrose2006} for an account of immunology's priority disputes in the 17\textsuperscript{th} and 18\textsuperscript{th} centuries).
Major impetus to study the lymphatic system came only in the 1990s, after the discovery of a suitable lymphatic marker that allowed quantifiable observation of lymphatic dynamics \cite{choi2012,oliver2002}. Lymphatic research is still a current trend in biomedicine and a source of sensational new discoveries, such as the 2015 finding of lymphatic vessels in the central nervous system \cite{louveau2015}.

An impetus for studying lymphatic regeneration is provided by recent biological studies that propose lymphangiogenesis as a major target for the treatment of non-healing wounds: functional lymphangiogenesis is nowadays regarded as a crucial factor in wound healing \cite{cho2006,ji2005,oliver2002,witte2001} and delayed or failed lymphatic regeneration (such as that observed in diabetic patients) constitutes a major cause of impairment to wound healing \cite{asai2012,maruyama2007,saaristo2006}.

Interest in lymphatics is therefore not just a mere scientific curiosity: their importance as pressure regulators in tissues and, moreover, as vectors of the immune response has been emphasised in recent decades, particularly in the context of wound healing \cite{cho2006,huggenberger2011,ji2005}.
The healing of a skin wound is a complex process consisting of different overlapping phases that, if well orchestrated by the organism, lead to the restoration of the skin and vasculature to a healthy, functional condition. Unfortunately, this delicate sequence of events can fail to proceed to full healing in diabetic or aged patients \cite{asai2012,jeffcoate2003,swift2001}; indeed, if the organism response to infection is insufficient, wound healing does not proceed through all normal stages, halting at the inflammation stage and resulting in a chronic wound \cite{brem2007,pierce2001}.

Non-healing wounds constitute a major health problem, seriously affecting the patient's quality of life and accounting for approximately 3\% of all health care expenses in the UK \cite{drew2007,posnett2008}. Being the main mediators of the immune response, lymphatics seem to significantly contribute to healing \cite{oliver2002,witte2001} and it has been observed that failed lymphangiogenesis correlates with impaired wound healing \cite{asai2012,maruyama2007,saaristo2006}.
However, little is known about the actual mechanisms involved in the lymphangiogenic process, in contrast with the (blood) angiogenic case \cite{benest2008,choi2012}.

Mathematical modelling potentially provides an alternative, powerful tool to back up experimental observations, generate a better understanding of wound healing lymphangiogenesis and identify potential clinical targets. Here we build upon our ODE model presented in \cite{bianchi2015} to address the spatial elements of lymphangiogenesis, specifically focussing on modelling two different hypotheses proposed to describe the exact lymphangiogenesis mechanism.

\section{Biological background}

\subsection{Wound healing}

For educational purposes, wound healing is usually presented as a sequence of four different (overlapping) phases, namely:
\begin{enumerate}
\item \emph{Hemostasis}: Shortly after injury, a blood clot is formed as a result of the interaction between blood and the extracellular matrix; the clot stops the bleeding and provides a scaffold for cells and chemicals that will consequently contribute to the healing process.
\item \emph{Inflammation}: Substances activated during hemostasis attract \emph{leukocytes}, inflammatory cells which clean the wound from debris and pathogens and secrete chemicals which promote the evolution of the system to the next phase.
\item  \emph{Proliferation}: The chemicals released during inflammation enhance the growth and aggregation of the surrounding cells, restoring different tissue functions and elements such as the blood and lymphatic networks; the regeneration of blood and lymphatic vessels is named \emph{(blood) angiogenesis} and \emph{lymphangiogenesis}, respectively. In this phase, the blood clot is slowly substituted by a ``temporary dermis'' called \emph{granulation tissue}. In parallel with these processes, the rapid proliferation and migration of epidermal cells causes this outer layer of the skin to re-form.
\item \emph{Remodelling}: Finally, the granulation tissue is slowly replaced by normal skin tissue; this stage can take up to two years to be completed.
\end{enumerate}
For further details about wound healing we refer to \cite{singer1999} for normal cutaneous wound healing, and to \cite{stadelmann1998} for an account of chronic wound dynamics.


\subsection{Sprouting versus self-organising lymphangiogenesis}  \label{sec:BioLymphang}

The lymphatic system is a circulatory system responsible for mediating the immune response of the body and maintaining the physiological pressure in tissues by draining excess liquid. It is mainly constituted of vessels and lymph nodes. Lymphatic vessel walls are made of so-called \emph{lymphatic endothelial cells} (LECs); contrary to the blood case, lymphatic capillaries are very thin and are formed of a single layer of LECs.

\begin{figure}[h]
     \includegraphics[width=0.8\textwidth]{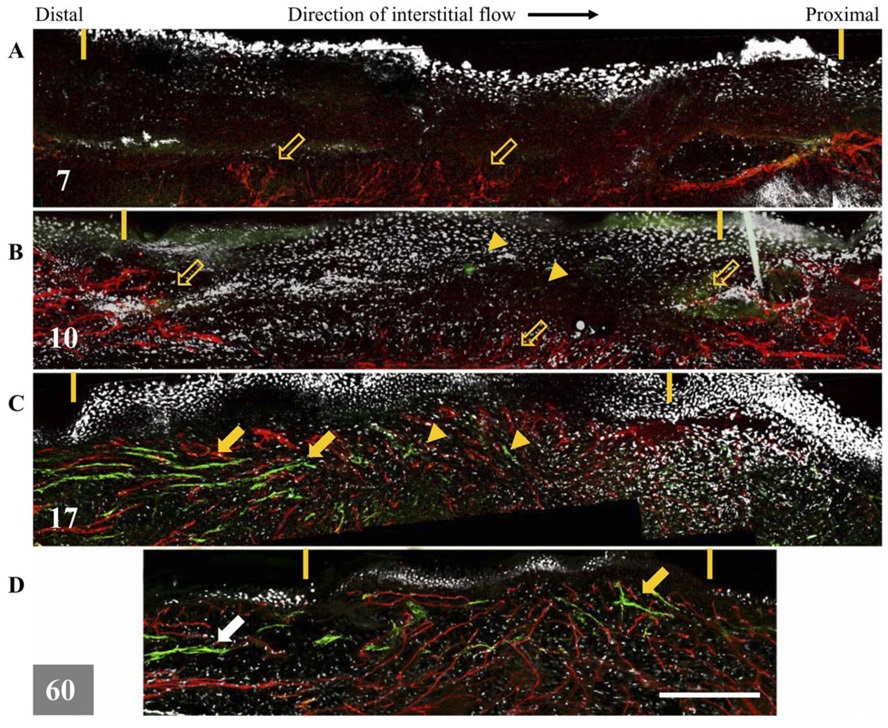}
		\caption{ In the photo, taken from \cite[Figure 2]{rutkowski2006}, one observes blood and lymphatic vessel regeneration in the tail of an adult mouse; lymphangiogenesis appears to occur in the direction of the interstitial flow. The different photos refer to different times after wounding: A was taken at day 7, B at day 10, C at day 17 and D at day 60. The yellow dashes mark the regenerating region (note its overall contraction over time); the red colour indicates blood vessels, while LEC presence is highlighted by green colour. The open arrows signal how blood vessels seem to sprout from deeper vessels, while other arrows point out LEC organisation at day 17 after a higher LEC density is reached; arrowheads indicate single LECs migrating towards the proximal side of the wound. Scale bar in D = 300 $\mu$m.   }
	\label{fig:LECrutkowski}
\end{figure}

To date, little is known about the biological events taking place during lymphangiogenesis and different hypotheses have been proposed by biologists. Although important reviews on the subject such as \cite{norrmen2011,tammela2010} state that lymphangiogenesis ``occurs primarily by sprouting from pre-existing vessels'', in a fashion which resembles the (blood) angiogenic case, recent experiments suggest that this may not be correct, at least not in some specific experimental settings \cite{benest2008,rutkowski2006}. In \cite{benest2008} it is stated that lymphangiogenesis ``can occur both by recruitment of isolated lymphatic islands to a connected network and by filopodial sprouting''. Similarly, in \cite{rutkowski2006} it is reported that in an adult mouse tail wound model, LECs migrate as single cells into the wound space and later connect to each other forming vessel structures (see Figure \ref{fig:LECrutkowski}). According to the authors of \cite{rutkowski2006}, single LEC migration following the lymph/interstitial flow would explain why lymphatic vessel regeneration appears to happen in this direction (from left to right in the figure).
Comparative reviews of lymphangiogenesis and (blood) angiogenesis can be found in \cite{adams2007,lohela2009,sweat2012}.

\subsection{Interstitial versus lymph flow}

Interstitial flow is a fluid flow induced by dynamic stresses and pressure gradients through the extracellular matrix. It is generally slower than fluid flow inside vessels, because of the resistance of the extracellular matrix components; nonetheless, interstitial flow has recently been shown to play an important role in many processes, including cell migration.  Such effects can be purely mechanical, for example by ``pushing'' on the cell, or can act indirectly by shifting the distribution of chemicals in the surroundings of the cell. 
A review of the effects of interstitial flow on cell biology can be found in \cite{rutkowski2007}.


In recent years, a number of studies have investigated the role of interstitial flow on lymphangiogenesis, mainly through the formation of concentration gradients of pro-lymphangiogenic factors. In particular, in \cite{boardman2003} the authors propose that interstitial flow, enhanced by the lymph flow resulting from interrupted lymphatic vessels, can direct wound healing lymphangiogenesis by transporting LECs into the wound space and creating gradients in chemicals (such as vascular endothelial growth factor -- VEGF) which stimulate LEC growth and chemotaxis.
However, the relative role of interstitial and lymph flow on capillary regeneration has yet to be investigated in depth; therefore, it is not clear which of the two takes on the greatest importance. In fact, although interstitial flow is slower than the flux of the lymph coming from interrupted capillaries, the former persists after wound closure, 
while the latter is more localised to open capillaries and stops once the lymphatic network has been restored.


\section{Mathematical Modelling}  

\subsection{Review of lymphatic-related models}

Contrary to the blood angiogenesis case, modelling literature about lymphangiogenesis is relatively immature and sparse, and mostly refers to tumour-induced lymphangiogenesis (see for instance \cite{friedman2005}). To the authors' knowledge, the only models addressing lymphangiogenesis in wound healing are \cite{roose2008}, which focuses on the mechanical elements that lend the lymphatic network its characteristic shape (at least in the mouse tail), and our previous work \cite{bianchi2015}, which we are going to extend here. A recent review of mathematical models of vascular network formation is \cite{scianna2013}, where indeed the imbalance between blood angio- and lymphangio-genesis modelling is manifest.

A number of models have been produced by the bioengineering community, describing specific mechanical features of lymphatic physiology; in particular, mechanics of contracting lymph valves have been presented in \cite{galie2009,heppell2015,macdonald2008,mendoza2003,reddy1995}.
A brief review of engineering models proposed in the lymphatic context can be found in \cite{margaris2012}.

Very few attempts have been made to specifically model the effect of flow on capillary regeneration, although one interesting example is \cite{fleury2006}, where the authors use a convection-diffusion model to analyse the effects of flow on matrix-binding protein gradients.

\subsection{Model targets}

The model hereby presented aims to investigate the following questions about wound healing lymphangiogenesis:
\begin{itemize}
\item which hypothesis (self-organising or sprouting) offers a better explanation for the lymphangiogenesis mechanics?
\item what are the relative contributions of interstitial and lymph flow on the lymphangiogenic process?
\item how does the initial wounded state impact on lymphatic regeneration?
\end{itemize}

\subsection{Model variables and domain}

In the following, we propose two similar but distinct PDE models to describe the two different theories advanced by biologists to explain lymphangiogenesis in wound healing (see Section \ref{sec:BioLymphang}). We will refer to them as the  ``self-organising'' hypothesis (O) and the  ``sprouting'' hypothesis (S). 

For both cases, we consider the following basic dynamics: immediately after injury, transforming growth factor-$\beta$ (TGF-$\beta$) is activated and chemotactically attracts macrophages to the wound, which in turn secrete VEGF which induces capillary regeneration acting on either LECs (in the self-organising case) or capillary tips (in the sprouting case).
The variables included in the models are summarised in Table \ref{tab:PDEvariables}, where they are reported together with their names and units.

\begin{table}[h]
\caption{A summary of the model variables.}
\label{tab:PDEvariables}

\begin{tabular}{cccc}
\hline\noalign{\smallskip}
\textsc{variable}  & \textsc{model}  &  \textsc{quantity}           &  \textsc{units}  \\
\noalign{\smallskip}\hline\noalign{\smallskip}
     $T(t,x)$      & O,S  & active TGF-$\beta$ concentration        & $\mbox{pg}\cdot\mbox{mm}^{-3}$     \\
     $M(t,x)$      & O,S  & macrophage density                      & $\mbox{cells}\cdot\mbox{mm}^{-3}$  \\
     $V(t,x)$      & O,S  & VEGF concentration                      & $\mbox{pg}\cdot\mbox{mm}^{-3}$     \\
     $L(t,x)$      & O    & lymphatic endothelial cell density      & $\mbox{cells}\cdot\mbox{mm}^{-3}$  \\
     $E(t,x)$      & S    & lymphatic capillary end (tip) density   & $\mbox{cells}\cdot\mbox{mm}^{-3}$  \\
     $C(t,x)$      & O,S  & lymphatic capillary density             & $\mbox{cells}\cdot\mbox{mm}^{-3}$  \\
\noalign{\smallskip}\hline
\end{tabular}

\end{table}

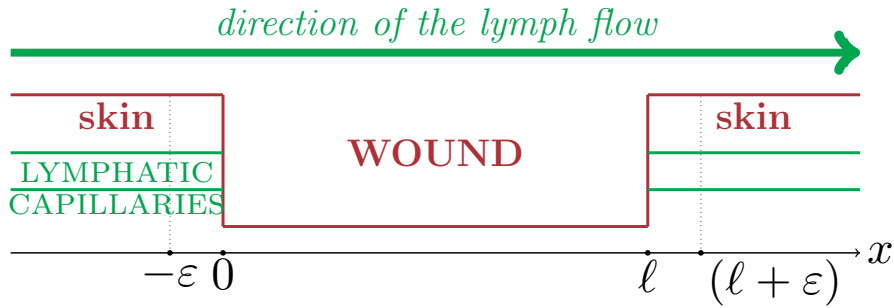
\begin{figure}[h]


\resizebox{0.7\textwidth}{!}{%

\begin{tikzpicture}
  \draw[->] [thick]  (-8,0) -- (8,0) node[right] {\scalebox{2.5}{$x$}};  
  \node [below, name=zero] at (-5,0) {\scalebox{2.5}{$-\varepsilon$}};
  \fill (zero.north) circle [radius=1.5pt];
  \node [below, name=epsil] at (-4,0) {\scalebox{2.5}{$0$}};
  \fill (epsil.north) circle [radius=1.5pt];
  \node [below, name=ell-epsil] at (4,0) {\scalebox{2.5}{$\ell$}};
  \fill (ell-epsil.north) circle [radius=1.5pt];
  \node [below right, name=ell] at (5,0) {\scalebox{2.5}{$(\ell+\varepsilon)$}};
  \fill (ell.north west) circle [radius=1.5pt];
  \draw[-] [dotted] (-5,0) -- (-5,3);
  \draw[-] [dotted] (5,0) -- (5,3);

  \draw[-] [ultra thick, Maroon]  (-8,3) -- node[below] {\textbf{\huge skin}} (-4,3) ;
  \draw[-] [ultra thick, Maroon]  (-4,3) -- (-4,0.5) ;
  \draw[-] [ultra thick, Maroon]  (-4,0.5) -- node[above=1cm] {\textbf{\huge WOUND}} (4,0.5) ;
  \draw[-] [ultra thick, Maroon]  (4,0.5) -- (4,3) ;
  \draw[-] [ultra thick, Maroon]  (4,3) -- node[below] {\textbf{\huge skin}} (8,3) ;

  \draw[->] [line width=0.15cm, Green, decoration={markings,mark=at position 1 with {\arrow[scale=1.5,Green]{>}}},
    postaction={decorate}, shorten >=0.4pt]  
    (-8,3.8) -- node[above] {\em{\huge direction of the lymph flow}} (8,3.8) ;
  \draw[-] [ultra thick, Green]  (-8,1.9) -- (-4,1.9) ;
  \draw[-] [ultra thick, Green]  (-8,1.2) -- node[above, name=lymph1] {{\fontsize{0.71cm}{0.5cm}\selectfont \textsc{lymphatic}}} (-4,1.2) ;
  \node [below=0.3cm, Green] at (lymph1) {{\fontsize{0.71cm}{0.5cm}\selectfont \textsc{capillaries}}};
  \draw[-] [ultra thick, Green]  (4,1.9) -- (8,1.9) ;
  \draw[-] [ultra thick, Green]  (4,1.2) -- (8,1.2) ;

\end{tikzpicture}

}

\caption{The model 1D domain.}

\label{fig:domain}

\end{figure}

We consider a 1D space variable $x$ that varies between $-\varepsilon$ and $\ell+\varepsilon$; this interval includes the wound space of length $\ell$ and a portion $\varepsilon$ of healthy tissue on its edges. 
This kind of domain describes a narrow cut, where at every point we average chemical and cell densities over the depth of the wound.
We take the increasing-$x$ direction to be that of lymph flow (and interstitial flow). A schematic of the model domain is shown in Figure \ref{fig:domain}.

\subsection{Advection velocity and open capillaries}
\label{sec:AdvectionCop-intro}

The models incorporate an advection term for the majority of variables that accounts for the effect of flow on the lymphatic regeneration process. In biological references (such as \cite{boardman2003}) it is not clear whether flow is mainly a result of lymph fluid exiting the interrupted capillaries, or the ``normal'' interstitial flow. We hence investigate the relative contribution from these two components by considering an advection term motivated as follows.

In general, interstitial flow does not have a constant direction. However, for simplicity, here we will assume that both lymph and interstitial flow occur in the increasing direction of $x$ (from left to right in Figure \ref{fig:domain}); this reflects what is observed in the wound healing experimental setting of \cite{boardman2003}, which we take as a reference for model comparison. We assume the interstitial flow to be constant and present across the full tissue, reflecting its persistent nature in healthy tissues. On the other hand, the contribution due to leaking lymphatic capillaries is assumed to depend specifically on the density of open capillaries $C_{op}$ and we assume a linear dependence for simplicity. However, since we do not know the precise contribution of each element to the total advection, we introduce a single  ``weight'' parameter $\xi$, $0\leq\xi\leq 1$, which can be varied. Specifically, the advection velocities for chemicals and cells, $\lambda^{chem}$ and $\lambda^{cell}$ respectively, will be taken to be of the forms
\begin{eqnarray}
\lambda^{chem}(C_{op}) & = & \xi \cdot ( \lambda_1^{chem} \cdot C_{op} ) + (1-\xi)\cdot\lambda_2^{chem} \; \mbox{ and }
 \label{eq:def-lambdaChem}  \\
\lambda^{cell}(C_{op}) & = & \xi \cdot ( \lambda_1^{cell} \cdot C_{op} ) + (1-\xi)\cdot\lambda_2^{cell} \; ,
 \label{eq:def-lambdaCell}
\end{eqnarray}
where $0\leq\xi\leq 1$ and $\lambda_1^{chem}$,$\lambda_2^{chem}$,$\lambda_1^{cell}$,$\lambda_2^{cell}$ are four parameters to be determined. In Appendix \ref{app-lambdas} we estimate the values of $\lambda_1^{chem}$ and $\lambda_2^{chem}$, while corresponding parameters for cells are assumed to be significantly smaller, since advective cell velocity is likely to be smaller due to the higher environmental friction.
A value of $\xi=0$ corresponds to purely interstitial flow advection, while $\xi=1$ represents advection due entirely to lymphatic flow.

\begin{figure}[h]

\resizebox{0.8\textwidth}{!}{%

\begin{tikzpicture}
  \draw[->] [thick] (0,0) -- (5,0) node[below right] {$x$};
  \draw[->] [thick] (0,0) node[below] {$0$} -- (0,3.5) ;
  \draw[thick,scale=1,domain=0:5,smooth,variable=\x,dashed,blue] plot ({\x},{ 1.5*( 1 - tanh( 1.5*(\x-2) ) )});
  \node [right,blue] at (0.5,3.3) {$C$};
  \draw[thick,scale=1,domain=0:5,smooth,variable=\x,red]
        plot ({\x},{ 1.5*( 1 - tanh( 1.5*(\x-2) ) )*2.25*( 1 - (tanh( 1.5*(\x-2) ))^2 )/(5+2.25*( 1 - (tanh( 1.5*(\x-2) ))^2 ))  });
  \node [right,red] at (0.5,0.75) {$C_{op}$};
\end{tikzpicture}
\quad
\begin{tikzpicture}
  \draw[->] [thick] (0,0) -- (5,0) node[below right] {$x$};
  \draw[->] [thick] (0,0) node[below] {$0$} -- (0,3.5) ;
  \draw[thick,scale=1,domain=0:5,smooth,variable=\x,dashed,blue] plot ({\x},{ 1.5*( 1 - tanh( 0.5*(\x-2) ) )});
  \node [right,blue] at (1,2.5) {$C$};
  \draw[thick,scale=1,domain=0:5,smooth,variable=\x,red]
        plot ({\x},{ 1.5*( 1 - tanh( 1.5*(\x-2) ) )*0.75*( 1 - (tanh( 1.5*(\x-2) ))^2 )/(5+0.75*( 1 - (tanh( 1.5*(\x-2) ))^2 ))  });
  \node [right,red] at (0.5,0.5) {$C_{op}$};
\end{tikzpicture}

}

\caption{Plots of $C_{op}$ ({\color{red}solid red}) for different steepness of $C$ ({\color{blue}dashed blue}).}

\label{fig:CopCplots}

\end{figure}
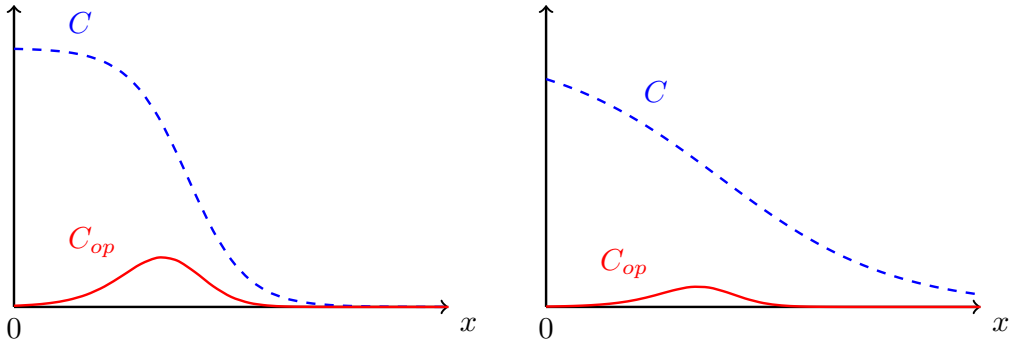

To quantify the open capillary density, we assume that as the ``cut'' in capillary density $C$ becomes steeper (and thus $|\nicefrac{\partial C}{\partial x}| \rightarrow +\infty$), more capillaries are open and the open capillary density will increase towards its maximum possible value of $C$, which would correspond to all capillaries being open. We therefore define the open capillary density $C_{op}$ as
\begin{equation}  \label{eq:CopenDef}
C_{op} \left( C,\frac{\partial C}{\partial x} \right) = \frac{|\nicefrac{\partial C}{\partial x}|}{\eta_0 + |\nicefrac{\partial C}{\partial x}|}\cdot C
\end{equation}
where $\eta_0$ is a parameter for whose estimation no relevant experimental data were found. See Figure \ref{fig:CopCplots} for a plot of \eqref{eq:CopenDef}.

\subsection{Self-organising hypothesis}

Under this hypothesis, single LECs migrate into the wound and start to self-organise into capillary structures only after reaching a certain threshold density $L^*$. This case represents the direct extension of the ODE model developed in \cite{bianchi2015} and the variable and parameter names have been kept as consistent with \cite{bianchi2015} as possible.


\subsubsection*{(Active) TGF-$\beta$ equation}

The differential equation describing active TGF-$\beta$ concentration has the following form:

\begin{center}
\resizebox{\textwidth}{!}{%
    \begin{tabular}{ C{3cm} c C{3cm} c c c c c C{3cm}}
  change in TGF-$\beta$ concentration & $=$ & diffusion and advection & $+$ &  activation & $-$ & decay & $-$ & internalisation by macrophages.
    \end{tabular}
}
\end{center}

\noindent
Of these terms, the following three are assumed to have standard forms:
$$
\mbox{Diffusion: } D_T \frac{\partial^2 T}{\partial x^2} \quad , \quad
\mbox{Decay: } d_1 T \quad , \quad
\mbox{Internalisation: } \gamma_1 T M \quad ,
$$
and advection will be taken to be $-\nicefrac{\partial }{\partial x}(\lambda^{chem}(C_{op})\cdot T)$, with velocity $\lambda^{chem}(C_{op})$ as defined in \eqref{eq:def-lambdaChem}.

Concerning the activation, we consider a constant amount of latent TGF-$\beta$ in the skin $T_L$ \cite{taylor2009,shi2011}, which is increased by macrophage production at rate $r_1$ \cite{khalil1993}.
This latent form of TGF-$\beta$ is activated by macrophages \cite{taylor2009,decrescenzo2001,gosiewska1999,nunes1995} and by the enzymes (mainly plasmin) present in the blood clot, which is mainly composed of platelets \cite{grainger1995,hyytiainen2004,khalil1996} (for a review of TGF-$\beta$ activation see \cite{taylor2009}).
Therefore, we take the following activation term:
\begin{equation*}
\underbrace{ \left[ a_m M + a_p p(C)\right] }
                             _{\stackrel{ \mbox{\footnotesize \footnotesize activation by macro-}}{ \mbox{\footnotesize -phages \& plasmin}}}
                \cdot \underbrace{ \left[ T_L + r_1 M \right] }_{ \mbox{\footnotesize latent TGF-}\beta} \; .
\end{equation*}
The $C$-dependent quantity $p$ is an estimate of plasmin presence in the wound, which is proportional to the platelet mass. In fact, although activation of platelet-released TGF-$\beta$ is still poorly understood, it seems that plasmin, while degrading the blood clot, activates the latent TGF-$\beta$ contained in the platelets \cite{grainger1995}.
We assume that the plasmin level is proportional to the wound space which is not occupied by capillaries; this is motivated by the fact that capillary presence can be considered as a measure of the healing stage of the wound.\footnote{An alternative approach would be to consider fibroblasts instead of capillaries here, but the introduction of a new variable and consequently a new equation does not seem to be worthwhile, since capillary presence is a good indication of the healing state of the wound.} When capillary density gets close to its equilibrium (healthy state) value $C^{eq}$ (say 90\% of it), the plasmin-induced TGF-$\beta$ activation switches to zero.
We will thus take
\begin{equation}  \label{eq:def-p(C)}
p(C) = \left\{ \begin{array}{cl}
-\frac{\psi}{\nicefrac{9}{10}\cdot C^{eq}} C + \psi & \mbox{if } C \leq ( C^{eq} \cdot \nicefrac{9}{10} ) \\
         0                 & \mbox{if } C \geq ( C^{eq} \cdot \nicefrac{9}{10} )    \; .
\end{array}  \right.
\end{equation}

%
%
%
%
%
%



\subsubsection*{Macrophage equation}

The following scheme will be considered for macrophage dynamics:

\begin{center}
\resizebox{\textwidth}{!}{%
    \begin{tabular}{ C{2.5cm} c C{3cm} c C{2.5cm} c C{1.7cm} c C{1.7cm} c C{1cm} c C{3.7cm} }
  change in macrophage density & $=$ & random movement and advection & $+$ & chemotaxis by TGF-$\beta$ & $+$ & constant source & $+$ & influx from open capillaries \\
    &  $-$ & removal and differentiation & $-$ & crowding effect. & & & & &
    \end{tabular}
}
\end{center}

\noindent
Macrophages are assumed to move randomly with diffusion coefficient $\mu_M$, while their advection will be modelled by the term $- \frac{\partial}{\partial x}\left( \lambda^{cell}(C_{op}) \cdot M \right)$, with $\lambda^{cell}(C_{op})$ as discussed in Section \ref{sec:AdvectionCop-intro}.

For the chemotaxis term, we first point out that only a fraction $\alpha$ of the monocytes that are chemoattracted by TGF-$\beta$ differentiate into (inflammatory) macrophages  \cite{mantovani2004,wahl1987}. Therefore, the term describing macrophage chemotaxis up TGF-$\beta$ gradients will have the form
$$
- \alpha \chi_1 \frac{\partial}{\partial x}\left( \frac{M}{1+\omega M}\cdot \frac{ \nicefrac{\partial T}{\partial x} }{ 1 + \eta_1 \left| \nicefrac{\partial T}{\partial x}\right| }  \right)
$$
where the macrophage velocity $\frac{1}{1+\omega M}\cdot \frac{ \nicefrac{\partial T}{\partial x} }{ 1 + \eta_1 \left| \nicefrac{\partial T}{\partial x}\right| }$ decreases as cell density increases (as in \cite{velazquez2004a,velazquez2004b}) and is bounded as $\left|\nicefrac{\partial T}{\partial x}\right|\rightarrow\infty$.
The presence of a constant source $s_M$ (from the bottom of the wound) is justified by the observation that the macrophage equilibrium in unwounded skin is nonzero \cite{weber1990}.

The introduction of an influx term is motivated by the fact that macrophages are ``pumped out'' from interrupted capillaries \cite{boardman2003,rutkowski2006} and into the wound. We consider the following form for the influx term:
\begin{equation}  \label{eq:MPh-InfluxTerm}
 \varphi_1 \left( C_{op} ,\frac{\partial C}{\partial x} \right) = C_{op} \cdot \zeta_1\left( \frac{\partial C}{\partial x} \right) \; ,
\end{equation}
where $C_{op}$ was introduced in \eqref{eq:CopenDef} and $\zeta_1$ is defined as
\begin{equation}  \label{eq:zeta1def}
\zeta_1\left( \frac{\partial C}{\partial x} \right) = \left\{ \begin{array}{cl}
                                                              \phi_1 & \mbox{if } \nicefrac{\partial C}{\partial x} <0 \\
                                                                 0   & \mbox{otherwise } \; .  \end{array}  \right.
\end{equation}
In \eqref{eq:zeta1def} $\phi_1$ is a parameter estimated in Appendix \ref{app-phi1param}.
The Heaviside form of $\zeta_1$ is due to the influx only occurring from the open lymphatic capillaries on the side of the wound from which lymph fluid flows (see Figure \ref{fig:domain}).

The removal term includes (inflammatory) macrophage death, differentiation into repair macrophages and reintroduction into the vascular system, with the latter being proportional to the capillary density. Thus, we take the removal term to be $( d_2 + \rho C ) M $.
We also include a crowding effect through the term $- \frac{M+L+C}{k_1} \cdot M$.



\subsubsection*{VEGF equation}

For VEGF we assume the following dynamics:

\begin{center}
\resizebox{\textwidth}{!}{%
    \begin{tabular}{ C{2.5cm} c C{3cm} c C{3cm} c C{2.2cm} c c}
  change in VEGF concentration & $=$ & diffusion and advection & $+$ & constant source & $+$ & production by macrophages \\
	 & $-$ & decay & $-$ & internalisation by LECs. & & & &
    \end{tabular}
}
\end{center}

\noindent
VEGF diffusion is modelled via the standard term $D_V \frac{\partial^2 V}{\partial x^2}$ and advection by $- \frac{\partial}{\partial x}\left( \lambda^{chem}(C_{op}) \cdot V \right)$ where $\lambda^{chem}(C_{op})$ is the expression defined in \eqref{eq:def-lambdaChem}.
The constant source is called $s_V$, while the production term will be $r_3 M$ and the decay $d_3 V$.
Internalisation is assumed to be linearly dependent on LEC density and the corresponding term will consequently be $\gamma_2 V L$.



\subsubsection*{LEC equation}

The equation describing the presence of LECs in the wound consists of the following terms:

\begin{center}
\resizebox{\textwidth}{!}{%
    \begin{tabular}{ C{2cm} c C{3cm} c C{2.5cm} c C{4cm} }
    change in LEC density & $=$ &  random movement and advection & $+$ & chemotaxis by VEGF & $+$ &growth, upregulated by VEGF and downregulated by TGF-$\beta$ \\
		& $+$ &  influx from open capillaries & $-$ & crowding effect & $-$ &  transdifferentiation into capillaries.
    \end{tabular}
}
\end{center}

\noindent
Again, random cell movement is modelled via a diffusion term $\mu_L \nicefrac{\partial^2 L}{\partial x^2}$ and the advection is taken to be $- \frac{\partial}{\partial x}\left( \lambda^{cell}(C) \cdot L \right)$.

LECs are chemoattracted by VEGF \cite{bernatchez1999,tammela2010}, and the chemotaxis term is assumed to be of a similar form to that used to describe macrophage chemotaxis:
$$
- \chi_2 \frac{\partial}{\partial x} \left( \frac{L}{1+\omega L} \cdot \frac{ \nicefrac{\partial V}{\partial x} }{ 1 + \eta_2\left| \nicefrac{\partial V}{\partial x}\right| } \right) \; .
$$

LEC growth is upregulated by VEGF \cite{bernatchez1999,whitehurst2006,zachary2001}
and downregulated by TGF-$\beta$ \cite{muller1987,sutton1991}:
$$
 \left( c_1 + \frac{V}{c_2 + c_3 V} \right) \left( \frac{1}{1+c_4 T} \right) L \; .
$$

LECs are ``pumped out'' from the interrupted capillaries in a similar manner  to macrophages, but also result (with less intensity) from interrupted capillaries downstream of the lymph flow.
The influx term this time takes the form:
\begin{equation} \label{eq:LEC-InfluxTerm}
\varphi_2 \left( C_{op} ,\frac{\partial C}{\partial x} \right) = C_{op} \cdot \zeta_2 \left( \frac{\partial C}{\partial x} \right)
\end{equation}
where $C_{op}$ is the density of open capillaries as in \eqref{eq:CopenDef} and $\zeta_2$ is defined as
\begin{equation}  \label{eq:zeta2def}
\zeta_2\left( \frac{\partial C}{\partial x} \right) = \left\{ \begin{array}{cl}
                                                      \phi_2^+ & \mbox{if } \nicefrac{\partial C}{\partial x} <0 \\
                                                      \phi_2^- & \mbox{if } \nicefrac{\partial C}{\partial x} >0  \; ,  \end{array}  \right.
\end{equation}
where $\phi_2^+>\phi_2^-$.

LECs cannot grow excessively due to crowding, which is taken into account via the term $-\frac{(M+L+C)}{k_2}\cdot L$.
When LECs have \emph{locally} sufficiently populated the wound (i.e. when their density exceeds a threshold $L^*$ \cite{boardman2003,rutkowski2006}) they are assumed to self-organise into capillaries at a rate which is increased by the presence of VEGF \cite{podgrabinska2002}:
$$
\sigma(L,C)\cdot (\delta_1 + \delta_2 V)L
$$
where
\begin{equation} \label{eq:sigmadef}
\sigma(L,C) = \left\{ \begin{array}{cl}
               1   &  \mbox{ if } L+C \geq L^*  \\
               0   &  \mbox{ if } L+C   <  L^*  \; .
\end{array}  \right.
\end{equation}


\subsubsection*{Lymphatic capillary equation}

After LECs have occupied enough of the wound space, they coalesce into a capillary network; also, they undergo remodelling, which we model via a logistic term. Thus, the $C$-equation will be
$$
\underbrace{\sigma(L,C)\cdot (\delta_1 + \delta_2 V)L}_{\mbox{source}}
+ \underbrace{ c_5 \left( 1 - \frac{C}{k_3}\right) C}_{ \mbox{remodelling} }  \; .
$$
Observe that no advection term is present here, since capillary structures are collections of cells attached to each other and thus are more resistant to the interstitial flows.


\subsubsection*{Full system -- ``self-organising'' hypothesis}

\begin{figure}[h]


\begin{tikzpicture}
  \draw[->] [thick] (-1,0) node[below] {$0$} -- (10,0) node[below right] {$x$};
  \draw[thick,scale=1,domain=0:9,smooth,variable=\x,black] plot ({\x},{ 4.24*( 1 - 0.5*(tanh( 1.1*(\x-0.4) ) + tanh( 1.1*(-\x+8.6) ) ) });
  \node [left] at (0,3) {$C$};
  \draw[->] [ultra thick,red] (-0.4,3.5)  -- node[below] {\footnotesize \textsc{interstitial flow}} (9.5,3.5)  ;
  \draw[->] [thick,blue] (0.3,1.6)  -- (2.2,1.6) node[below] {\footnotesize \textsc{macrophages}} ;
  \draw[->] [ultra thick,red] (0,2.5)  -- (1.9,2.5) node[below] {\footnotesize \textsc{lymph}} ;
  \draw[->] [ultra thick,red] (7.9,2.5)  -- (9.8,2.5) node[below] {\footnotesize \textsc{lymph}} ;
  \draw[->] [thick,green] (0.5,0.8)  -- (2.4,0.8) node[below] {\footnotesize \textsc{LECs}} ;
  \draw[->] [thick,green] (8.5,0.8)  -- (7.4,0.8) node[below left] {\footnotesize \textsc{LECs}} ;
\end{tikzpicture}

\caption{A summary of the fluxes included in the model: \textbf{capillaries}; {\color{red}\textbf{fluid fluxes}}; {\color{blue}\textbf{macrophage influx}};  {\color{green}\textbf{LEC influx} (only in O)}.}

\label{fig:FluxesSummary}

\end{figure}
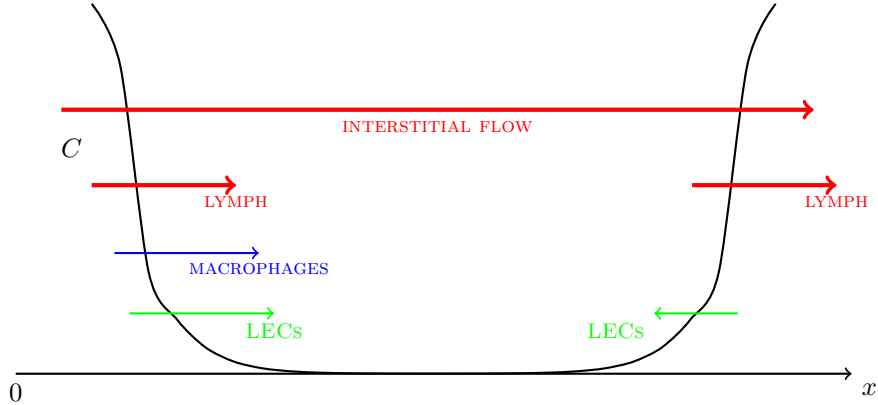

The full system of equations in the ``self-organising'' hypothesis is therefore given by

\begin{eqnarray}
\frac{\partial T}{\partial t} & = & D_T \frac{\partial^2 T}{\partial x^2} - \frac{\partial}{\partial x}\left( \lambda^{chem}(C_{op}) \cdot T \right)
+ \left[ a_m M + a_p p(C)\right] \cdot \left[ T_L + r_1 M \right] \nonumber \\
 & & - d_1 T - \gamma_1 T M \; , \label{eq:SelfOrg-T} \\
\frac{\partial M}{\partial t} & = & \mu_M \frac{\partial^2 M}{\partial x^2} - \frac{\partial}{\partial x}\left( \lambda^{cell}(C_{op}) \cdot M + \alpha \chi_1 \frac{M}{1+\omega M}\cdot \frac{ \nicefrac{\partial T}{\partial x} }{ 1 + \eta_1\left| \nicefrac{\partial T}{\partial x}\right| }  \right) \nonumber \\
& &  + s_M + \varphi_1 \left( C_{op} , \frac{\partial C}{\partial x} \right)  - ( d_2 + \rho C ) M  - \frac{M+L+C}{k_1}M  \; ,  \label{eq:SelfOrg-M} \\
\frac{\partial V}{\partial t} & = & D_V \frac{\partial^2 V}{\partial x^2} - \frac{\partial}{\partial x}\left( \lambda^{chem}(C_{op}) \cdot V \right) + s_V + r_3 M - d_3 V - \gamma_2 VL  \; , \label{eq:SelfOrg-V} \\
\frac{\partial L}{\partial t} & = & \mu_L \frac{\partial^2 L}{\partial x^2} - \frac{\partial}{\partial x}\left( \lambda^{cell}(C_{op}) \cdot L + \chi_2 \frac{L}{1+\omega L} \cdot \frac{ \nicefrac{\partial V}{\partial x} }{ 1 + \eta_2\left| \nicefrac{\partial V}{\partial x}\right| } \right)  \nonumber \\
 & & + \left( c_1 + \frac{V}{c_2 + c_3 V} \right) \left( \frac{1}{1+c_4 T} \right) L + \varphi_2 \left( C_{op} , \frac{\partial C}{\partial x} \right)    \nonumber \\
                              &   &  - \frac{M+L+C}{k_2}L - \sigma(L,C)\cdot (\delta_1 + \delta_2 V)L  \; , \label{eq:SelfOrg-L} \\
\frac{\partial C}{\partial t} & = & \sigma(L,C)\cdot (\delta_1 + \delta_2 V)L + c_5 \left( 1 - \frac{C}{k_3}\right) C  \; ,  \label{eq:SelfOrg-C}
\end{eqnarray}
where $\lambda^{chem}$ is defined in \eqref{eq:def-lambdaChem}, $\lambda^{cell}$ in \eqref{eq:def-lambdaCell}, $p$ in \eqref{eq:def-p(C)}, $\varphi_1$ in \eqref{eq:MPh-InfluxTerm}, $\varphi_2$ in \eqref{eq:LEC-InfluxTerm} and $\sigma$ in \eqref{eq:sigmadef}.
Parameters, initial and boundary conditions are discussed in Sections \ref{sec:PDEpar} and \ref{sec:PDEICandBCs} respectively.
See Figure \ref{fig:FluxesSummary} for a summary of the fluxes included in the model.

\subsection{Sprouting hypothesis}

Here, instead of LECs we consider capillary tip density $E$. Capillary tips are attached to the vessel ends and therefore, contrary to LECs, are not subject to advection. As we will see, the introduction of this variable is necessary in order to model directed capillary growth in response to a gradient. Examples of mathematical models of blood angiogenesis (in wound healing and in tumours) which include the capillary tip variable can be found in \cite{byrne1995,byrne2000,flegg2012,flegg2015,levine2001,mantzaris2004,schugart2008}.

TGF-$\beta$, macrophage and VEGF equations are the same as in the self-organising case, except that in both the crowding term for $M$ and the $V$ internalisation term there is $E$ instead of $L$.

\subsubsection*{Lymphatic capillary ends (tips) equation}
Capillary ends (or tips) are assumed to sprout from interrupted lymphatic capillaries, the density of which ($C_{op}$) was defined in \eqref{eq:CopenDef}. Tip growth is enhanced by VEGF and inhibited by TGF-$\beta$ and this is reflected by the following term, similar to the one used for LECs in the self-organising case:
$$
\left( c_1 + \frac{V}{c_2 + c_3 V} \right) \left( \frac{1}{1+c_4 T} \right) C_{op}  \; .
$$
Importantly, capillary ends move in the direction of the (positive) gradient of VEGF with an upper-bounded velocity, modelled by the term
$$
- \chi_2 \frac{\partial}{\partial x} \left( E \cdot \frac{ \nicefrac{\partial V}{\partial x} }{ 1 + \eta_2\left| \nicefrac{\partial V}{\partial x}\right| } \right) \; .
$$

Finally, we assume that capillary tip death is due predominantly to overcrowding, and thus we include the removal term $-\frac{(M+E+C)}{k_2}\cdot E$.

\subsubsection*{Lymphatic capillary equation}
New capillaries are formed continuously from the interrupted ones in the direction defined by their tips. This is modelled here according to the ``snail trail'' concept which was introduced in \cite{edelstein1982} for fungal colonies and which has been widely used in models of (blood) angiogenesis \cite{flegg2012}: newly formed capillaries are laid after the sprouting tips, which therefore leave a sort of ``track'' behind.

Capillaries also  undergo remodelling. Therefore, their dynamics are captured by the terms:
$$
\underbrace{\chi_2 \left| E \cdot \frac{ \nicefrac{\partial V}{\partial x} }{ 1 + \eta_2\left| \nicefrac{\partial V}{\partial x}\right| }  \right|}_{\mbox{sprouting}}
+ \underbrace{ c_5 \left( 1 - \frac{C}{k_3}\right) C}_{ \mbox{remodelling} }  \; .
$$

\subsubsection*{Full system -- ``sprouting'' hypothesis}

Thus, the full system for the ``sprouting'' hypothesis is

\begin{eqnarray}
\frac{\partial T}{\partial t} & = & D_T \frac{\partial^2 T}{\partial x^2}-\frac{\partial}{\partial x}\left( \lambda^{chem}(C_{op}) \cdot T \right)  + \left[ a_m M + a_p p(C)\right] \cdot \left[ T_L + r_1 M \right] \nonumber \\
& & - d_1 T - \gamma_1 T M  \; , \label{eq:Sprout-T} \\
\frac{\partial M}{\partial t} & = & \mu_M \frac{\partial^2 M}{\partial x^2} -\frac{\partial}{\partial x}\left( \lambda^{cell}(C_{op}) \cdot M + \alpha \chi_1 \frac{M}{1+\omega M}\cdot \frac{ \nicefrac{\partial T}{\partial x} }{ 1 + \eta_1\left|\nicefrac{\partial T}{\partial x}\right|}  \right)  \nonumber \\
& & + s_M  +  \varphi_1 \left( C_{op} , \frac{\partial C}{\partial x} \right)   - ( d_2 + \rho C ) M - \frac{M+E+C}{k_1}M  \; , \label{eq:Sprout-M} \\
\frac{\partial V}{\partial t} & = & D_V \frac{\partial^2 V}{\partial x^2} -\frac{\partial}{\partial x}\left( \lambda^{chem}(C_{op}) \cdot V \right) + s_V + r_3 M - d_3 V - \gamma_2 VE   \; , \label{eq:Sprout-V} \\
\frac{\partial E}{\partial t} & = & \left( c_1 + \frac{V}{c_2 + c_3 V} \right) \left( \frac{1}{1+c_4 T} \right) C_{op}
                                    - \chi_2 \frac{\partial}{\partial x} \left( E \cdot \frac{ \nicefrac{\partial V}{\partial x} }{ 1 + \eta_2\left| \nicefrac{\partial V}{\partial x}\right| } \right) \nonumber \\
																		& & - \frac{M+E+C}{k_2} E  \; , \label{eq:Sprout-E}  \\
\frac{\partial C}{\partial t} & = & \chi_2 \left| E \cdot \frac{ \nicefrac{\partial V}{\partial x} }{ 1 + \eta_2\left| \nicefrac{\partial V}{\partial x}\right| } \right|  +  c_5 \left( 1 - \frac{C}{k_3}\right) C    \; , \label{eq:Sprout-C}
\end{eqnarray}
where $\lambda^{chem}$ is defined in \eqref{eq:def-lambdaChem}, $\lambda^{cell}$ in \eqref{eq:def-lambdaCell}, $p$ in \eqref{eq:def-p(C)}, $\varphi_1$ in \eqref{eq:MPh-InfluxTerm} and $C_{op}$ in \eqref{eq:CopenDef} (see Figure \ref{fig:FluxesSummary} for a summary of the fluxes of the model).


\subsection{Parameters}  \label{sec:PDEpar}

All the model parameters are reported in Table \ref{table:param}. Many of the parameters were estimated previously in \cite{bianchi2015} and we refer to this source for details of their estimation.
For the other parameters listed in Table \ref{table:param}, the details of their estimation can be found in Appendix \ref{appPARpde}.

\begin{table}[p]

\begin{small}

\caption{ A list of parameters appearing in the model equations; those referred to \cite{bianchi2015} for details are the same as in the ODE model therein presented, while estimation of the newly introduced ones is discussed in Appendix \ref{appPARpde}. Each parameter is supplied with its estimated value, units and source used (when possible) to assess it. References in brackets mean that although the parameter was not \emph{directly} estimated from a dataset, its calculated value was compared with the biological literature; the caption ``no data found'' signifies that no suitable data were found to estimate the parameter. 
Note that $a_m$ here corresponds to $a_M$ in \cite{bianchi2015} and $\gamma_2$ here to $\gamma$ in \cite{bianchi2015}.
$k_1^{old}$ denotes the parameter $k_1$ in \cite{bianchi2015}, where it is the macrophage carrying capacity.
The parameter $d_4$ appears in the boundary conditions for $L$.
}

\label{table:param}

\makebox[\textwidth][c]{\begin{tabular}{ccccc}
\hline
\textsc{parameter}  &  \textsc{value} &  \textsc{units}  &  \textsc{source}  &  \textsc{details} \\
\hline
 $D_T$          &       2.76        & $\mbox{mm}^2\mbox{day}^{-1}$               & \cite{lee2014,murphy2012} & Appendix \ref{appPARpde} \\
 $\eta_0$       &       $10^4$          &   $\mbox{cells mm}^{-4}$                & no data found & Appendix \ref{appPARpde} \\
  $\lambda_1^{chem}$ & $1.35\times 10^{-2}$ & $\mbox{mm}\mbox{day}^{-1}$ & \cite{fischer1996,fischer1997} & Appendix \ref{appPARpde} \\
$\lambda_2^{chem}$ & $8.64\times 10^2$ & $\mbox{mm}\mbox{day}^{-1}$ & \cite{rutkowski2007} & Appendix \ref{appPARpde} \\
     $a_p$  & $2.9\times 10^{-2}$  & $\mbox{mm}^3\mbox{pg}^{-1}\mbox{day}^{-1}$  & \cite{decrescenzo2001} & \cite{bianchi2015} \\
      $\psi$         &      $10^5$       & $\mbox{pg mm}^{-3}$                        & no data found & Appendix \ref{appPARpde}\\
     $a_m$  & 0.45            & $\mbox{mm}^3\mbox{cells}^{-1}\mbox{day}^{-1}$   & \cite{gosiewska1999,nunes1995} & \cite{bianchi2015}  \\
     $T_L$  &  18              & $\mbox{pg mm}^{-3}$                             & (\cite{oi2004}) & \cite{bianchi2015}  \\
     $r_1$  & $3\times 10^{-5}$    & $\mbox{pg cells}^{-1}\mbox{day}^{-1}$      & \cite{khalil1993}  & \cite{bianchi2015}  \\
     $d_1$  & $5\times 10^2$ & $\mbox{day}^{-1}$                          & \cite{kaminska2005} & \cite{bianchi2015}   \\
    $\gamma_1$     & $4.2\times 10^{-3}$   & $\mbox{mm}^{3}\mbox{cells}^{-1}\mbox{day}^{-1}$ & (\cite{yang1999}) & Appendix \ref{appPARpde} \\
\hline
$\mu_M$        &      $0.12$       & $\mbox{mm}^2\mbox{day}^{-1}$               & \cite{farrell1990} & Appendix \ref{appPARpde} \\
$\lambda_1^{cell}$ & $1.35\times 10^{-3}$ & $\mbox{mm}\mbox{day}^{-1}$ & estimated $\approx 0.1\times\lambda_1^{chem}$ & Appendix \ref{appPARpde} \\
$\lambda_2^{cell}$ & $86.4$ & $\mbox{mm}\mbox{day}^{-1}$ & estimated $\approx 0.1\times\lambda_2^{chem}$ & Appendix \ref{appPARpde} \\
$\alpha$  & 0.5             & 1                                                     & \cite{waugh2006}  & \cite{bianchi2015}  \\
$\chi_1$       & $4\times 10^{-2}$ & $\mbox{mm}^5\mbox{pg}^{-1}\mbox{day}^{-1}$ & \cite{lijeon2002} & Appendix \ref{appPARpde} \\
$\omega$       & $1.67\times 10^{-6}$  & $\mbox{mm}^{3}\mbox{cells}^{-1}$       & estimated $\approx 1/k_1^{old}$  & Appendix \ref{appPARpde} \\
$\eta_1$       &       100          & $\mbox{mm}^9\mbox{pg}^{-1}$                & no data found   & Appendix \ref{appPARpde} \\
$s_M$          & $8.6\times 10^2$  & $\mbox{cells mm}^{-3}\mbox{day}^{-1}$   & (\cite{weber1990}) & Appendix \ref{appPARpde} \\
$\phi_1$       & $2.05\times 10^3$ & $\mbox{day}^{-1}$ & \cite{cao2005,fischer1996}  & Appendix \ref{appPARpde} \\
$\beta$  & $5\times 10^{-3}$           & 1                                                     & \cite{greenwood1973}  & \cite{bianchi2015}  \\
$r_2$    & 1.22          & $\mbox{day}^{-1}$                                     & \cite{zhuang1997}  & \cite{bianchi2015}  \\
$d_2$    & 0.2             & $\mbox{day}^{-1}$                             & \cite{cobbold2000} & \cite{bianchi2015}  \\
$\rho$   & $10^{-5}$     & $\mbox{day}^{-1}\mbox{cells}^{-1}$            & \cite{rutkowski2006} & \cite{bianchi2015}  \\
$k_1$          &       $10^5$      & $\mbox{mm}^{3}\mbox{cells}^{-1}$               & \cite{zhuang1997}  & Appendix \ref{appPARpde}  \\
\hline
 $D_V$          &       2.4         & $\mbox{mm}^2\mbox{day}^{-1}$               & \cite{miura2009}  & Appendix \ref{appPARpde} \\
     $s_V$  & 1.94 & $\mbox{cells}\mbox{ day}^{-1}$ & (\cite{hormbrey2003,papaioannou2009})  & \cite{bianchi2015}  \\
     $r_3$  & $1.9\times 10^{-3}$ & $\mbox{pg cells}^{-1}\mbox{day}^{-1}$  & (\cite{kiriakidis2003,sheikh2000}) & \cite{bianchi2015} \\
     $d_3$  & 11                & $\mbox{day}^{-1}$                              & \cite{kleinheinz2010}  & \cite{bianchi2015}  \\
   $\gamma_2$ & $1.4\times 10^{-3}$ & $\mbox{mm}^{3}\mbox{cells}^{-1}\mbox{day}^{-1}$& \cite{gabhann2004}  & \cite{bianchi2015} \\
\hline
 $\mu_L$        &       0.1         & $\mbox{mm}^2\mbox{day}^{-1}$               & estimated $\approx \mu_M$ & Appendix \ref{appPARpde} \\
     $c_1$          & 0.42         & $\mbox{day}^{-1}$                                     & \cite{nguyen2007}  & \cite{bianchi2015}  \\
     $c_2$          & 42         & $\mbox{day}$                                          & \cite{whitehurst2006} & \cite{bianchi2015} \\
     $c_3$          & 4.1        & $\mbox{pg day mm}^{-3}$                               & \cite{whitehurst2006} & \cite{bianchi2015}  \\
     $c_4$          & 0.24       & $\mbox{mm}^{3}\mbox{pg}^{-1}$                         & \cite{muller1987} & \cite{bianchi2015} \\
     $\chi_2$       &      0.173        & $\mbox{mm}^5\mbox{pg}^{-1}\mbox{day}^{-1}$ & \cite{barkefors2008}  & Appendix \ref{appPARpde} \\
 $\eta_2$       &       1           & $\mbox{mm}^9\mbox{pg}^{-1}$             & no data found & Appendix \ref{appPARpde}  \\
$\phi_2^+$     &     $10^2$              & $\mbox{day}^{-1}$                 & no data found  & Appendix \ref{appPARpde}  \\
 $\phi_2^-$     &       1             & $\mbox{day}^{-1}$                     & estimated to be 1\% of $\phi_2^+$ & Appendix \ref{appPARpde}  \\
     $k_2$          & $4.71\times 10^5$    & $\mbox{cells day mm}^{-3}$                     & \cite{nguyen2007} & \cite{bianchi2015} \\
     $L^*$          & $10^4$          & $\mbox{cells mm}^{-3}$                                & \cite{rutkowski2006}  & \cite{bianchi2015} \\
     $\delta_1$     & $5\times 10^{-2}$ & $\mbox{day}^{-1}$                                     & no data found  & \cite{bianchi2015} \\
     $\delta_2$     & $10^{-3}$         & $\mbox{mm}^3\mbox{pg}^{-1}\mbox{day}^{-1}$            & no data found  & \cite{bianchi2015} \\
\hline
$c_5$           &     0.42          & $\mbox{day}^{-1}$                   &  estimated = $c_1$ & Appendix \ref{appPARpde} \\
$k_3$           & $1.2\times 10^4$  & $\mbox{mm}^{3}\mbox{cells}^{-1}$    &  estimated $\approx C^{eq}$ & Appendix \ref{appPARpde}  \\
\hline
\end{tabular}}

\end{small}

\end{table}

\subsection{Initial and boundary conditions} \label{sec:PDEICandBCs}

\subsubsection*{Initial Conditions}

\begin{figure}[h]
\resizebox{0.95\textwidth}{!}{%
\begin{tikzpicture}
  \draw[->] [thick] (0,0) -- (10,0) node[right] {\scalebox{1.5}{$x$}};
  \draw[->] [thick] (0,0) node[below] {\scalebox{1.5}{$-\varepsilon$}} -- (0,4) ;
  \draw[-] [thick] (9,3.8) -- (9,0) node[below] {\scalebox{1.5}{$\ell+\varepsilon$}};
  \draw[-] [dashed,thick] (0.6,2.7) -- (0.6,0) node[below] {\scalebox{1.5}{$0$}};
  \draw[-] [dashed,thick] (8.4,2.7) -- (8.4,0) node[below] {\scalebox{1.5}{$\ell$}};
  \draw[thick,scale=1,domain=0:9,smooth,variable=\x,red] plot ({\x},{ 5*( 1 - 0.5*(tanh( 0.5*(\x-0.4) ) + tanh( 0.5*(-\x+8.6) ) )});
  \node [left,red] at (0,3) {\scalebox{2}{$T^{eq}$}};
  \node [below] at (4.5,2.5) {\textsc{\Large shallow wound}};
  \node [below] at (4.5,2.2) {\scalebox{2}{(e.g. $b=5$)}};
\end{tikzpicture}
\quad
\begin{tikzpicture}
  \draw[->] [thick] (0,0) -- (10,0) node[right] {\scalebox{1.5}{$x$}};
  \draw[->] [thick] (0,0) node[below] {\scalebox{1.5}{$-\varepsilon$}} -- (0,4) ;
  \draw[-] [thick] (9,3.8) -- (9,0) node[below] {\scalebox{1.5}{$\ell+\varepsilon$}};
  \draw[-] [dashed,thick] (0.6,2.7) -- (0.6,0) node[below] {\scalebox{1.5}{$0$}};
  \draw[-] [dashed,thick] (8.4,2.7) -- (8.4,0) node[below] {\scalebox{1.5}{$\ell$}};
  \draw[thick,scale=1,domain=0:9,smooth,variable=\x,red] plot ({\x},{ 4.24*( 1 - 0.5*(tanh( 1.1*(\x-0.4) ) + tanh( 1.1*(-\x+8.6) ) ) });
  \node [left,red] at (0,3) {\scalebox{2}{$T^{eq}$}};
  \node [below] at (4.5,2.5) {\textsc{\Large deep wound}};
  \node [below] at (4.5,2.2) {\scalebox{2}{(e.g. $b=100$)}};
\end{tikzpicture}
}

\caption{Initial condition $T(0,x)=a_T \cdot \{ 1 - [\tanh( b (x-\varepsilon) ) ) + \tanh( b(-x+\ell-\varepsilon)) ]/2 \}$ for different values of $b$.
$T^{eq}$ denotes the $T$-equilibrium level in non-wounded skin.}

\label{fig:ICplot}

\end{figure}
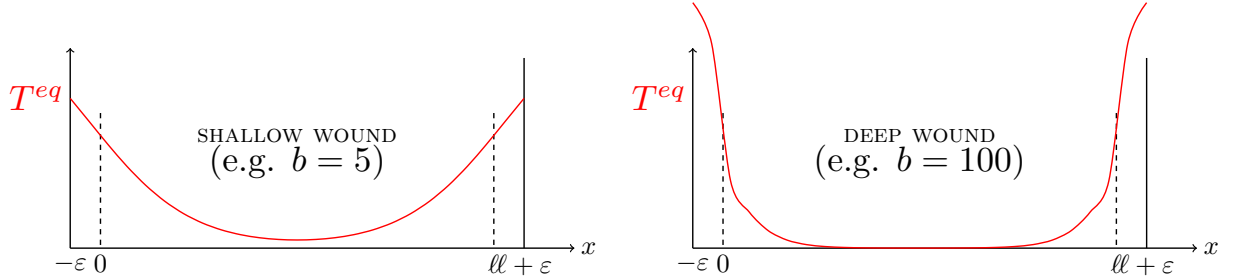

As initial time $t=0$ we take the moment of wounding, when little chemical or cell populations are assumed to have entered in the wound space.
Specifically, we assume that at $t=0$ there are no LECs (for model O) or capillary tips (for S), while other variables can be present near the edges (recall our domain includes portions of healthy skin surrounding the wound). We will then take the following initial conditions:
\begin{eqnarray}
\label{eq:PDEinitcondsTMVC}
\nu(0,x) & = & a_\nu \cdot
        \left[ 1 - \frac{\tanh( b x ) + \tanh( b(-x+\ell))}{2} \right] \; , \\
\label{eq:PDEinitcondsL+E}
L(0,x) & = & E(0,x) = 0 \; ,
\end{eqnarray}
where $\nu\in\{ T, M , V , C \} $.
For each variable $\nu$ the value of $a_{\nu}$ is chosen to be such that $\nu(0,-\varepsilon)=\nu(0,\ell+\varepsilon)$ is equal to the boundary conditions discussed in the following. Concerning $b$, we will vary its value to see how the ``sharpness'' of the initial condition will affect lymphangiogenesis. For higher values of $b$, the initial conditions become more step-like and we can interpret this as a deep wound with sharp edges: in this case, there would be (almost) no capillaries in the centre of the wound. On the other hand, assigning smaller values of $b$ would correspond to a shallower initial wound, such that when averaging over the wound depth a certain number of capillaries still remain.
As an example, the plot of \eqref{eq:PDEinitcondsTMVC} for $\nu =T$ is shown in Figure \ref{fig:ICplot} for different values of $b$.

\subsubsection*{Boundary Conditions}

First of all, note that boundary conditions are not needed for $C$. We consider Dirichlet boundary conditions for all other variables except $L$, for which we assume Robin boundary conditions. The choice of Dirichlet boundary conditions is dictated by the fact that at the boundary the tissue is in a non-wounded state, and we expect variables to remain close to their normal, equilibrium value there. For $L$, we apply instead the following reasoning.

For LECs, we assume that once they pass the domain edge they move randomly and die at a constant rate $d_4$; in fact, it seems unrealistic to assume that they will just vanish once reaching the domain edge. Therefore we will follow common practice for representation of habitat boundaries in ecological modelling \cite{ludwig1979}: we set a different evolution equation for $L$ inside and outside the domain. In the interior (i.e. for $-\varepsilon<x<\ell+\varepsilon$), the dynamics of $L$ will be described by the equation \eqref{eq:SelfOrg-L}; in the exterior (i.e. for $x<-\varepsilon$ and $x>\ell+\varepsilon$) instead we assume that LECs move randomly and die (or transdifferentiate) with (high) constant rate $d_4$. This gives the equation
\begin{equation}
\frac{\partial L}{\partial t} = \mu_L \frac{\partial^2 L}{\partial x^2} - d_4 L
\end{equation}
outside the wound, whose solution at equilibrium is given by
\begin{equation}
L_o (x) = A_o \exp\left( \sqrt{\frac{d_4}{\mu_L}}x \right) + B_o \exp\left( -\sqrt{\frac{d_4}{\mu_L}}x \right)
\end{equation}
where $A_o$ and $B_o$ are constants. Note that, since we want solutions to be bounded in order to be biologically meaningful, we will take $B_o=0$ for $x<-\varepsilon$ and $A_o=0$ for $x>\ell+\varepsilon$.
Since at the boundaries the outside and the inside solutions should have the same value and the same flux, we have that
\begin{eqnarray*}
& \mbox{at } x = -\varepsilon :    & \quad L = A_o \; \mbox{ and } \; \frac{\partial L}{\partial x} = A_o \sqrt{\frac{d_4}{\mu_L}} \quad \Rightarrow \; \frac{\partial L}{\partial x}(t,-\varepsilon) = \sqrt{\frac{d_4}{\mu_L}} L(t,0) \\
& \mbox{at } x = \ell+\varepsilon : & \quad L = B_o \exp\left( -\sqrt{\frac{d_4}{\mu_L}}\ell \right)    \; \mbox{ and } \; \frac{\partial L}{\partial x} = - B_o \sqrt{\frac{d_4}{\mu_L}}\exp\left( -\sqrt{\frac{d_4}{\mu_L}}\ell \right) \quad
               \\
               & & \Rightarrow \; \frac{\partial L}{\partial x}(t,\ell+\varepsilon) = -\sqrt{\frac{d_4}{\mu_L}} L(t,\ell)
\end{eqnarray*}
which give the boundary conditions for $L$.

Summarising, the boundary conditions are
\begin{eqnarray}
\nu (t,-\varepsilon) = \nu (t,\ell+\varepsilon) = \nu^{eq} \label{eq:PDEboundcondsTMVC} & , & E (t,-\varepsilon) = E (t,\ell+\varepsilon) = 0  \: , \\
\frac{\partial L}{\partial x} - \sqrt{\frac{d_4}{\mu_L}}L = 0 \quad \mbox{ at } x=-\varepsilon  & , &
\frac{\partial L}{\partial x} + \sqrt{\frac{d_4}{\mu_L}}L = 0 \quad \mbox{ at } x=\ell+\varepsilon  \label{eq:PDEboundcondsL}
\end{eqnarray}
with $\nu \in \{ T,M,V,C \} $ and where $\nu^{eq}$ denotes the equilibrium value in the unwounded skin for each variable.


\section{Numerical solutions} 

To simulate the two systems \eqref{eq:SelfOrg-T}-\eqref{eq:SelfOrg-C} and \eqref{eq:Sprout-T}-\eqref{eq:Sprout-C}, a specific code was written which applies the Crank-Nicolson method for the diffusion terms and a first-order upwind scheme for the chemotactic terms.

This section is structured as following: first, in \ref{sec:DataCmp}, we present the data sets which will be used as reference points in estimating the ``goodness'' of the simulations; then, in \ref{sec:firstSimOS} we present a sample simulation of both the whole O and S models; in \ref{sec:varyingbxi} we explore how changes in $b$ (initial condition steepness) and $\xi$ (interstitial/lymph flow balance) affect lymphatic regeneration; in \ref{sec:NoAdvAddAdv} we address the two extreme cases where there is no advection at all and where the two advection terms sum up (\emph{additive advection}); finally, in \ref{sec:OverallCmpr} we summarise all the observations concerning the different behaviour of O and S systems.

\subsection{Data for comparison} \label{sec:DataCmp}

\begin{figure}[ht]
\resizebox{\textwidth}{!}{%
\quad  \includegraphics[width=0.5\textwidth]{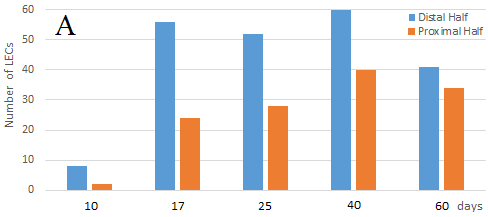}
\quad \includegraphics[width=0.5\textwidth]{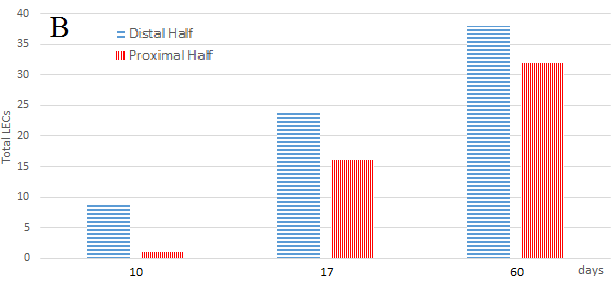} }
\caption{ Quantification of LEC presence and distribution in the regenerating region of a mouse tail wound. Here the total numbers of LECs in the distal and proximal halves of the wound at different days post-wounding are reported after data from (A) \cite[Figure 2]{rutkowski2006} and (B) \cite[Figure 1]{goldman2007}. }
\label{fig:LECdata}
\end{figure}

We will compare our model simulations with experimental data reported in Figure \ref{fig:LECdata}. These experimental observations show that the overall levels of LECs (both free and in a capillary structure) increase steadily after wounding, and that while at day 10 the vast majority are in the distal half (i.e. upstream the lymph flow) by 60 days they are almost evenly distributed over the two sides.

Hence, from experimental data:
\begin{itemize}
\item lymphatics should have reached a density close to $C^{eq}$ at day 60;
\item LEC migration and/or lymphatic capillary formation should happen predominantly in the direction of the lymph/interstitial flow.
\end{itemize}


\subsection{A first simulation of O and S} \label{sec:firstSimOS}

\begin{figure}[p]
\centering
\includegraphics[width=0.9\textwidth]{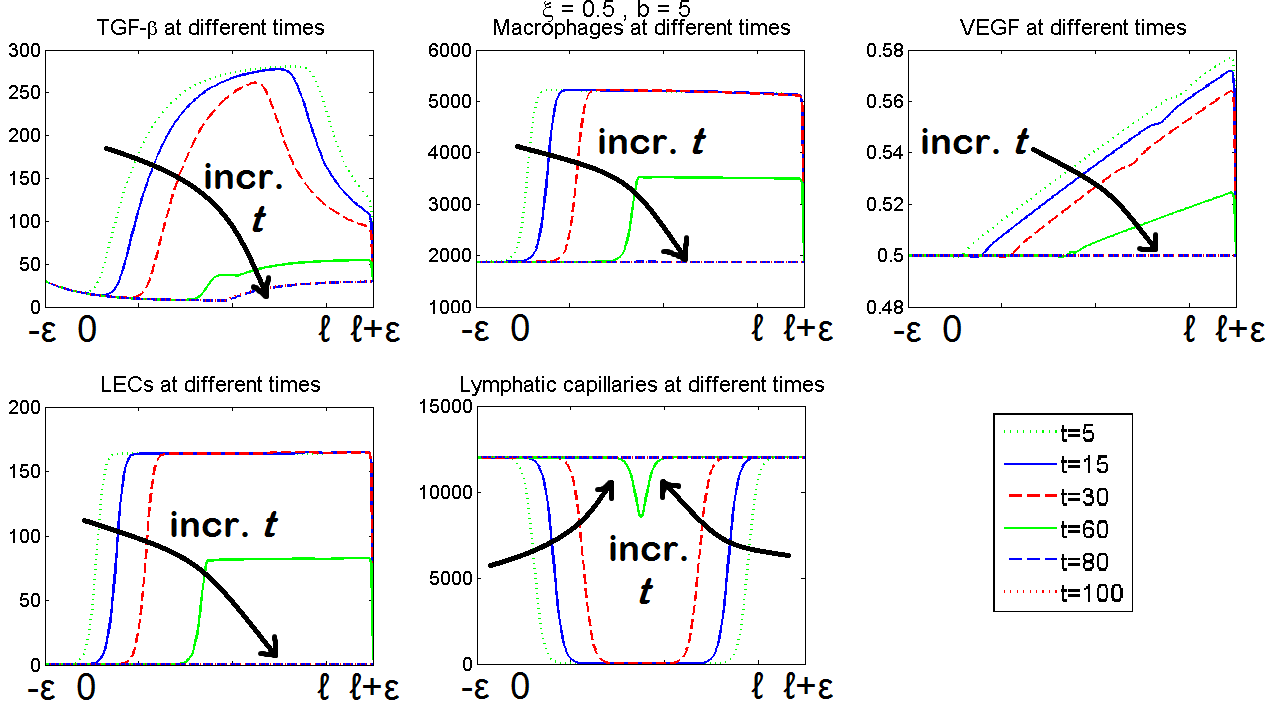}
\caption{Simulation of equations \eqref{eq:SelfOrg-T}--\eqref{eq:SelfOrg-C} (self-organising case) with parameters from Table \ref{table:param} and initial condition as defined in \ref{sec:PDEICandBCs}, with $b=5$; $\xi=0.5$.
Arrows mark the direction of increasing $t$ in the simulations.}
\label{fig:simSelfOrg-xi05tanh-b5}
\end{figure}
\begin{figure}[p]
\centering
\includegraphics[width=0.9\textwidth]{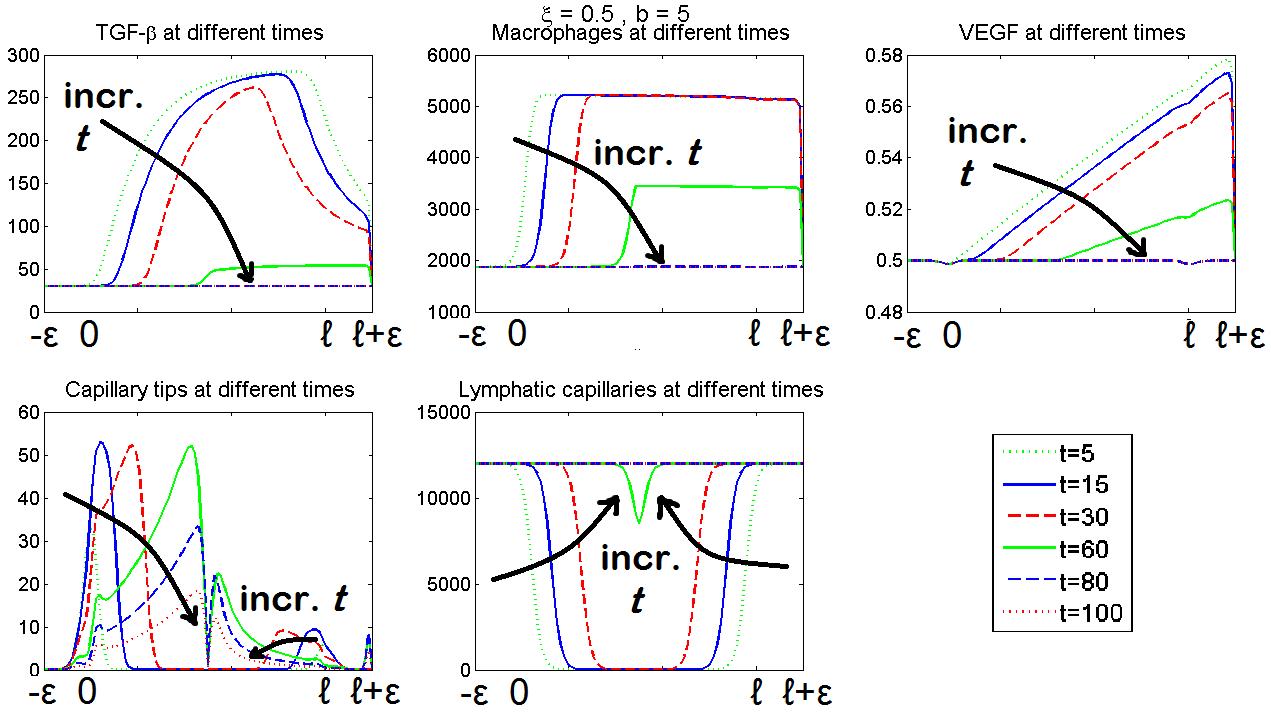}
\caption{Simulation of equations \eqref{eq:Sprout-T}--\eqref{eq:Sprout-C} (sprouting case) with parameters from Table \ref{table:param} and initial condition as defined in \ref{sec:PDEICandBCs}, with $b=5$; $\xi=0.5$.
Arrows mark the direction of increasing $t$ in the simulations.}
\label{fig:simSprout-xi05tanh-b5}
\end{figure}

We start by presenting simulations of the self-organising and sprouting cases (Figures \ref{fig:simSelfOrg-xi05tanh-b5} and \ref{fig:simSprout-xi05tanh-b5}, respectively)  with $\xi=0.5$ (representing that interstitial and lymph flow are equally weighted in the overall advection term) and a very smooth initial condition, with $b=5$ (see \eqref{eq:PDEinitcondsTMVC}).

For these values of $\xi$ and $b$, both systems predict lymphatic regeneration to be almost symmetric and a nearly-complete network is restored by around day 60 (see Figures \ref{fig:simSelfOrg-xi05tanh-b5} and \ref{fig:simSprout-xi05tanh-b5}). Biologically, this represents the situation in which a relatively shallow wound leaves more capillaries in the domain after wounding, so that regeneration occurs mainly from remodelling of the pre-existing network.
We note, however, that the distribution of the other variables is highly asymmetric. This will lead to a non-symmetric lymphatic regeneration when parameters are changed so that the chemical concentrations contribute more prominently to the lymphangiogenesis process.
One unexpected feature emerging from Figures \ref{fig:simSelfOrg-xi05tanh-b5} and \ref{fig:simSprout-xi05tanh-b5} is that macrophage, VEGF and LEC levels are higher than equilibrium in the healthy tissue on the right-hand-side of the wound, downstream the lymph flow. While some overspill is likely to be observed, particularly macrophage density appears to be too high to be realistic. In section \ref{sec:varyingbxi} we will present results suggesting that the value $\xi=0.5$ used in Figures \ref{fig:simSelfOrg-xi05tanh-b5} and \ref{fig:simSprout-xi05tanh-b5} is inappropriately low; the high downstream densities are a consequence of this. However an additional possible explanation might be that more processes are involved in bringing cell and chemical levels back to normal in the healthy skin surrounding a wound; macrophages are likely to be ``re-absorbed'' in the blood and lymphatic vasculature, where their number is balanced by factors not included in the model. However, the simulations shown in Figures \ref{fig:simSelfOrg-xi05tanh-b5} and \ref{fig:simSprout-xi05tanh-b5} do predict that eventually all the variables' amounts go back to equilibrium as healing proceeds.

\subsection{Varying $b$ and $\xi$} \label{sec:varyingbxi}

\paragraph{``Visual'' observations}

In order to clearly visualise the changes in dynamics when the parameters $b$ and $\xi$ are varied, we report the approximate solution profiles of the lymphatic capillary density at different times for different combinations of these two parameters; such simulations are reported in Tables \ref{tab:cfrOcapill} and \ref{tab:cfrScapill} for the self-organising and the sprouting case, respectively.


\begin{table}[p]

\caption{Plots of capillary density at different times for different values of $b$ and $\xi$ in the \textbf{self-organising case} (equations \eqref{eq:SelfOrg-T}--\eqref{eq:SelfOrg-C}); arrows mark the direction of increasing $t$ in the simulations.
On the right-hand-side of each box, we show bar plots of LEC presence (calculated as $L+C$) in distal (D) and proximal (P) half of the wound at days 10, 15, 25, 40 and 60 for different values of $b$ and $\xi$. }

\label{tab:cfrOcapill}

\resizebox{\textwidth}{!}{%
\begin{tabular}{ m{0.5cm} |c|c|}
\hline
\scalebox{4.5}{$\xi$} &  \scalebox{4.5}{ \textsc{shallow wound} ($b=5$) }  & \scalebox{4.5}{ \textsc{deep wound} ($b=100$) } \\
\hline
\raisebox{9\normalbaselineskip}[0pt][0pt]{\rot{\scalebox{3}{$\xi=0$}}}
        & \includegraphics[width=\linewidth]{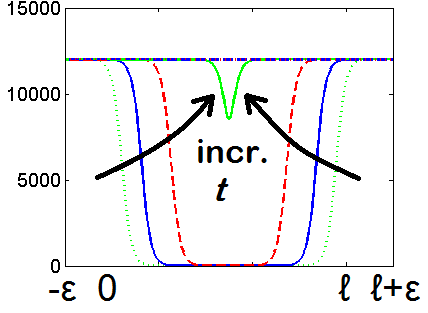}
          \includegraphics[width=1.5\linewidth]{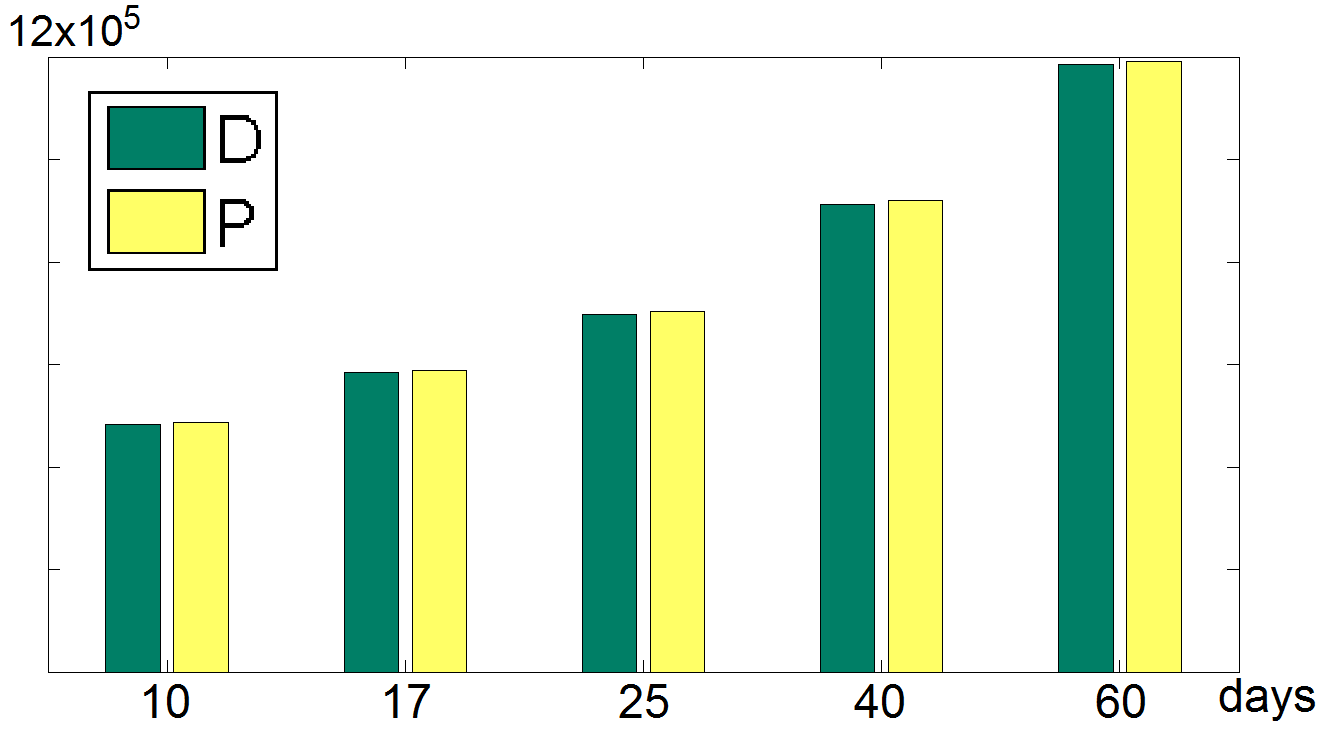}
        &    \includegraphics[width=\linewidth]{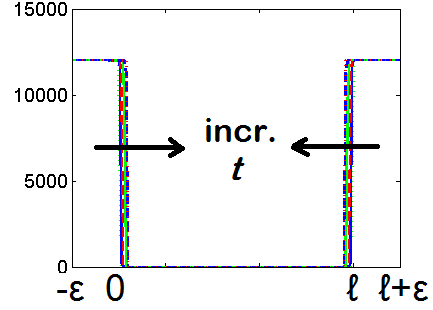}
             \includegraphics[width=1.5\linewidth]{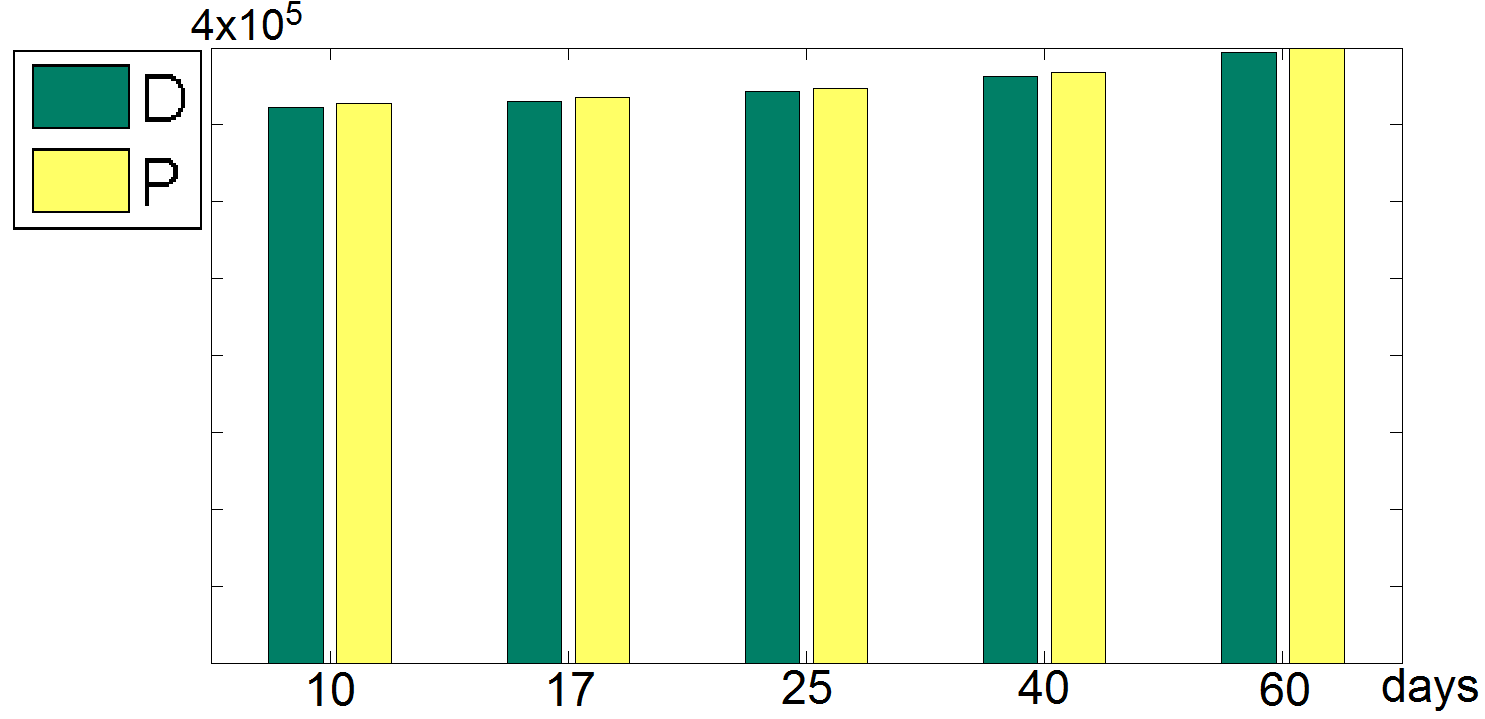}    \\
\hline
\raisebox{9\normalbaselineskip}[0pt][0pt]{\rot{\scalebox{3}{$\xi=0.75$}}}
        & \includegraphics[width=\linewidth]{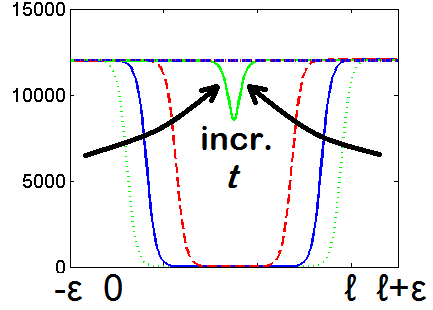}
          \includegraphics[width=1.5\linewidth]{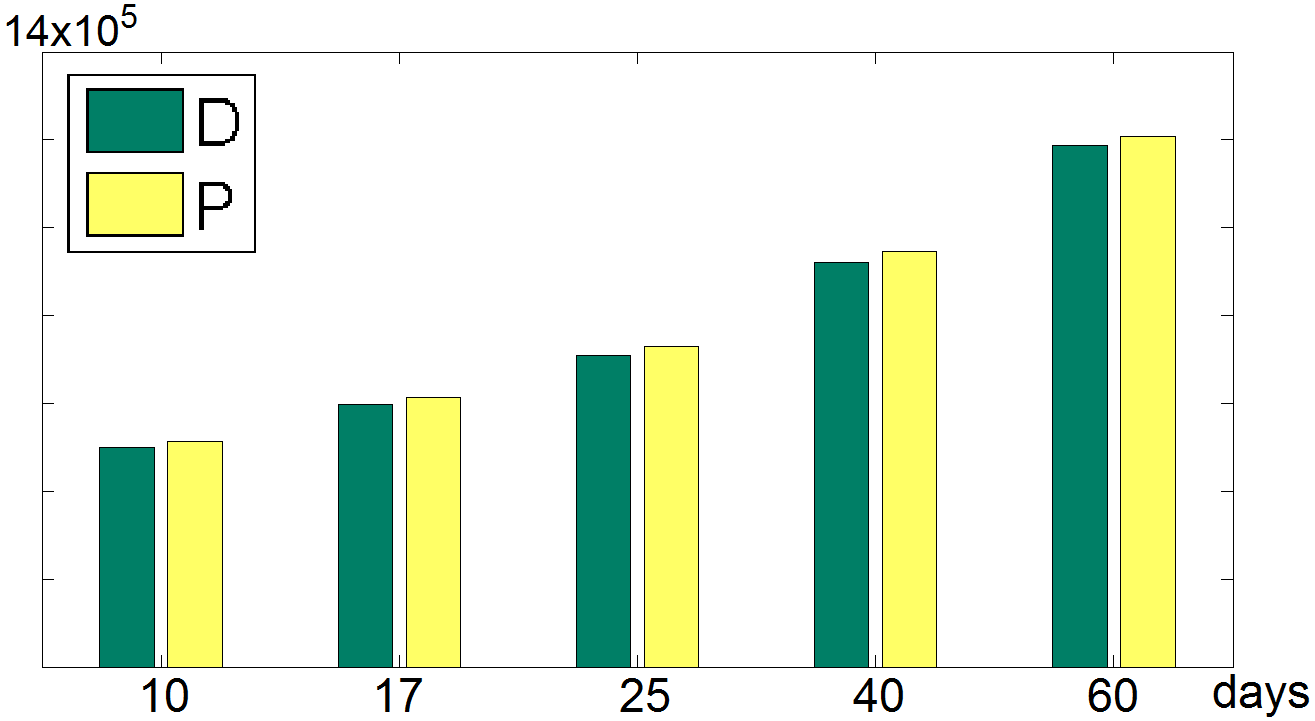}
        &  \includegraphics[width=\linewidth]{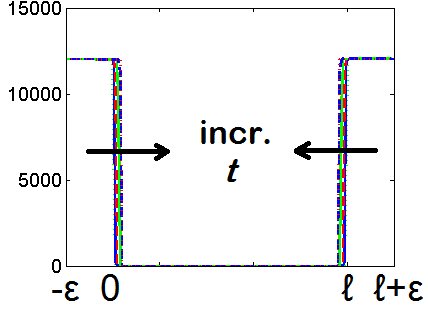}
           \includegraphics[width=1.5\linewidth]{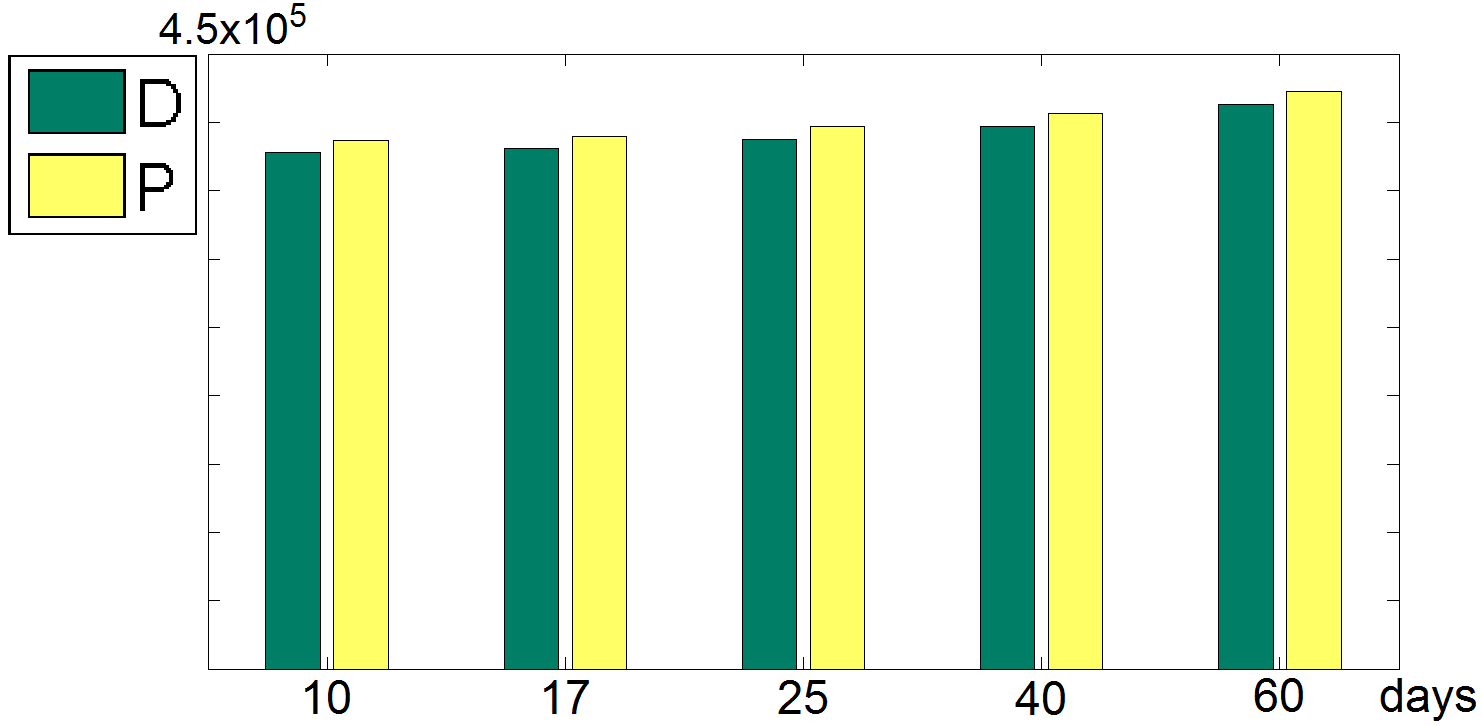}   \\
\hline
\raisebox{9\normalbaselineskip}[0pt][0pt]{\rot{\scalebox{3}{$\xi=1$}}}
        & \includegraphics[width=\linewidth]{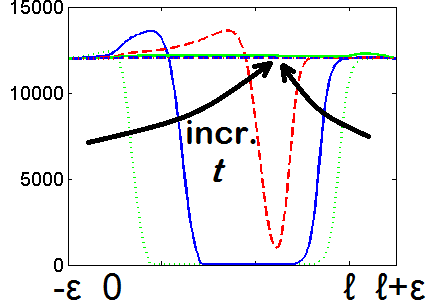}
          \includegraphics[width=1.5\linewidth]{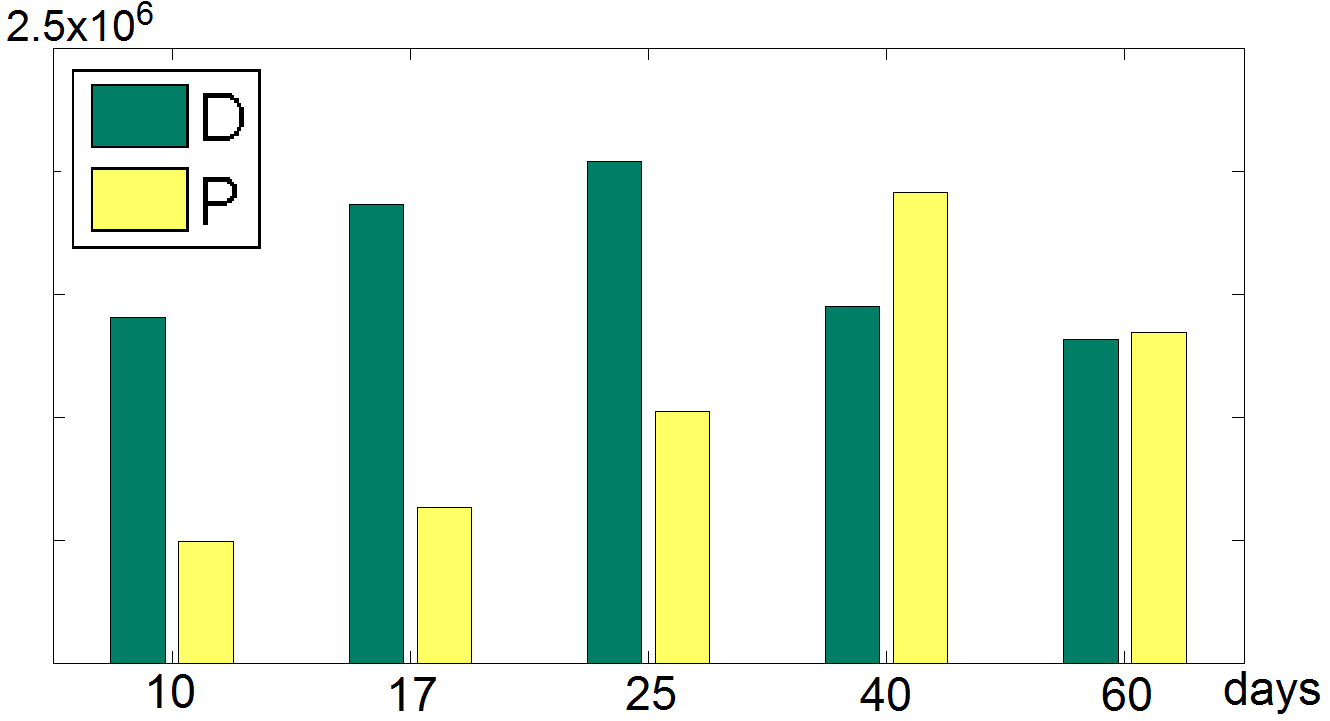}
        & \includegraphics[width=\linewidth]{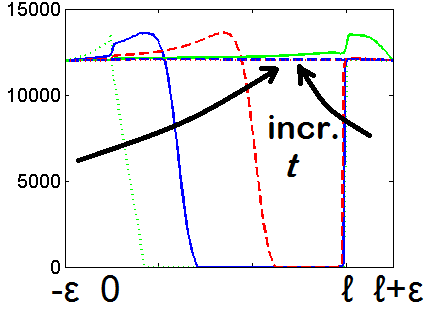}
          \includegraphics[width=1.5\linewidth]{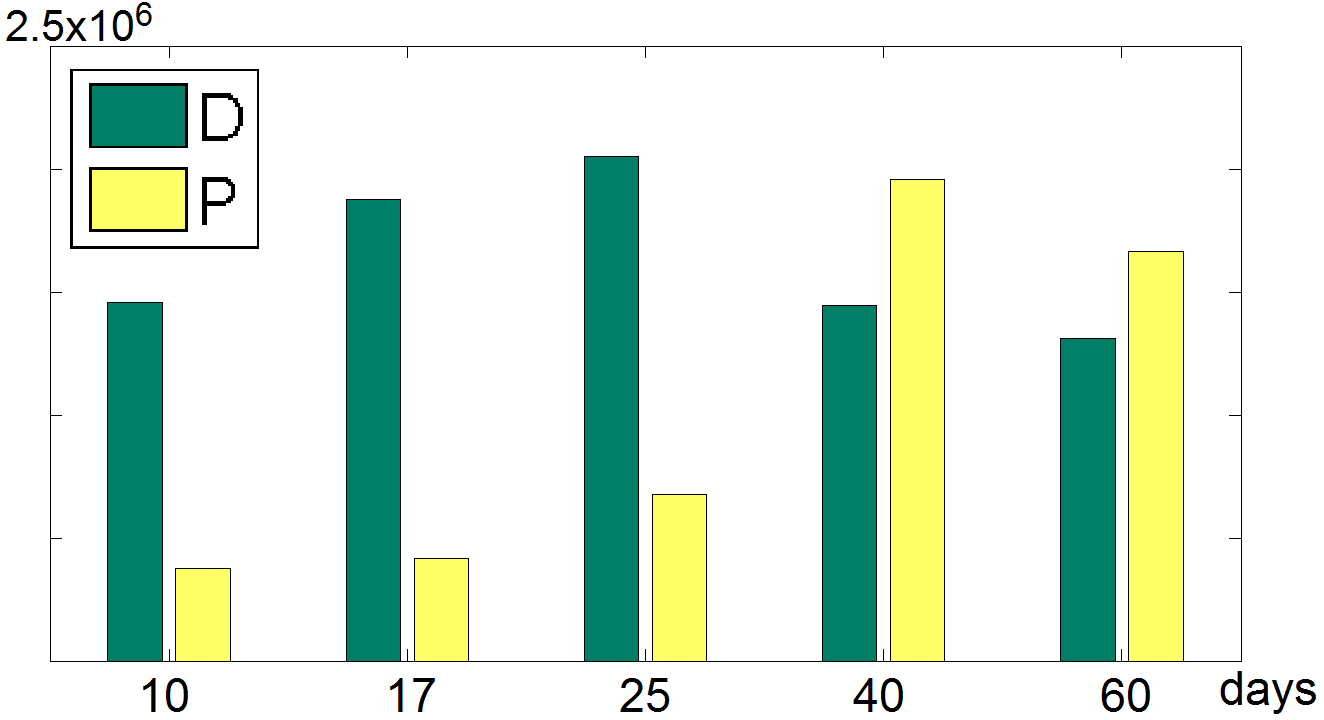}      \\
\hline
\end{tabular}
}

\end{table}


\begin{table}[p]

\caption{Plots of capillary density at different times for different values of $b$ and $\xi$ in the \textbf{sprouting case} (equations \eqref{eq:Sprout-T}--\eqref{eq:Sprout-C}); arrows mark the direction of increasing $t$ in the simulations.
On the right-hand-side of each box, we show bar plots of LEC presence (calculated as $E+C$) in distal (D) and proximal (P) half of the wound at days 10, 15, 25, 40 and 60 for different values of $b$ and $\xi$. }

\label{tab:cfrScapill}

\resizebox{\textwidth}{!}{%
\begin{tabular}{ m{0.5cm} |c|c|}
\hline
\scalebox{4.5}{$\xi$} & \scalebox{4.5}{ \textsc{shallow wound} ($b=5$)}  &  \scalebox{4.5}{ \textsc{deep wound} ($b=100$)} \\
\hline
\raisebox{9\normalbaselineskip}[0pt][0pt]{\rot{\scalebox{3}{$\xi=0$}}}
        & \includegraphics[width=\linewidth]{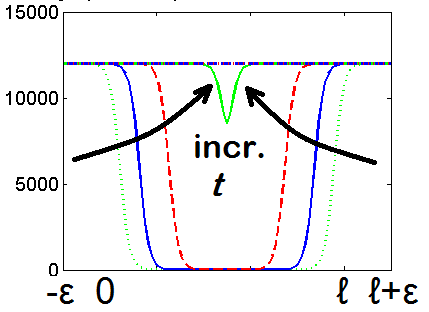}
        \includegraphics[width=1.5\linewidth]{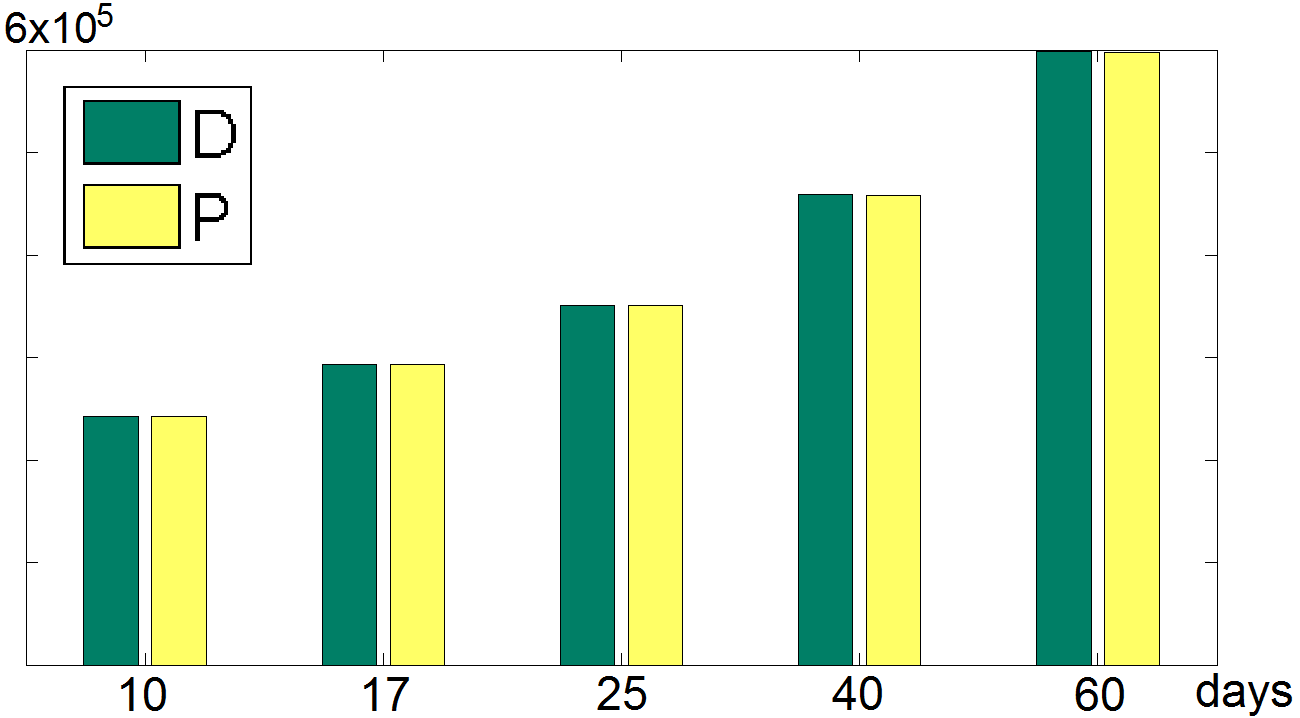}
        &  \includegraphics[width=\linewidth]{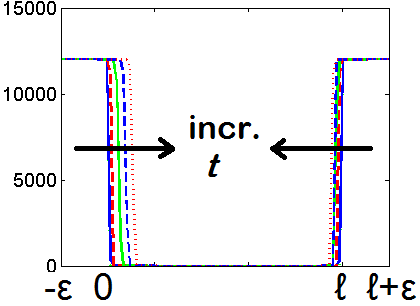}
         \includegraphics[width=1.5\linewidth]{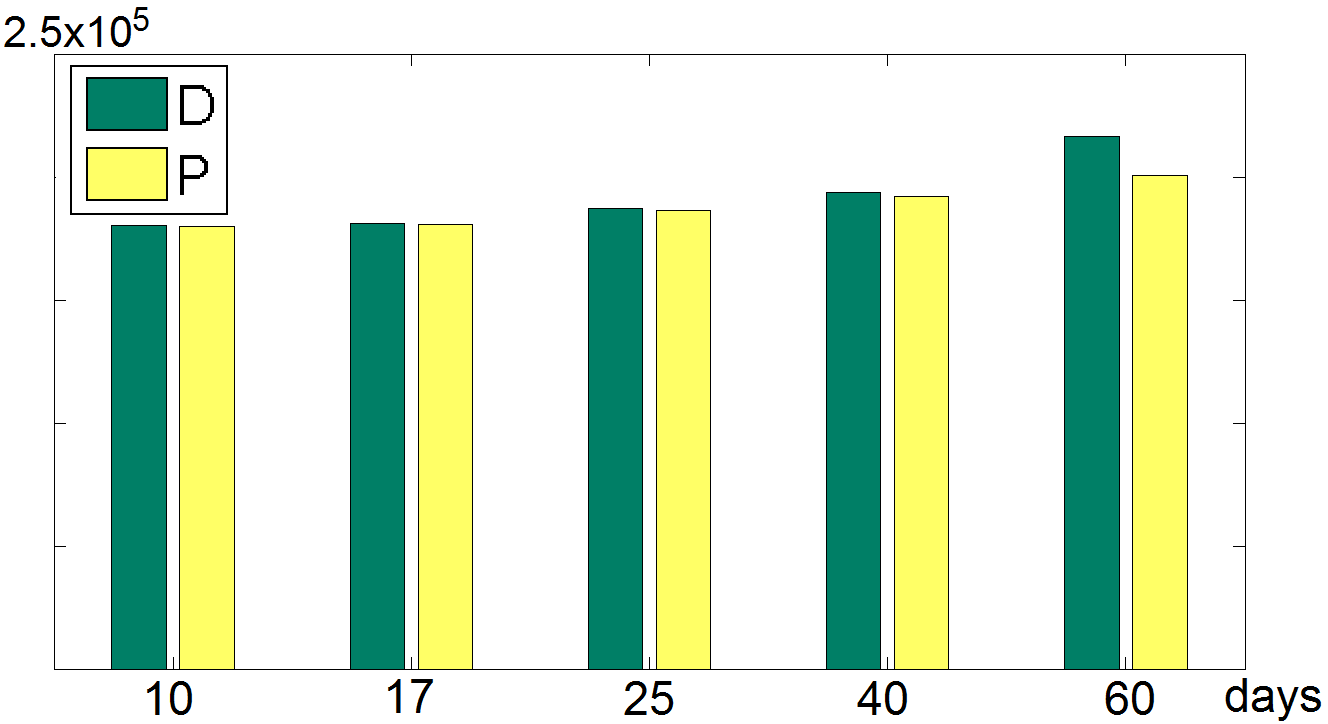}    \\
\hline
\raisebox{9\normalbaselineskip}[0pt][0pt]{\rot{\scalebox{3}{$\xi=0.75$}}}
        & \includegraphics[width=\linewidth]{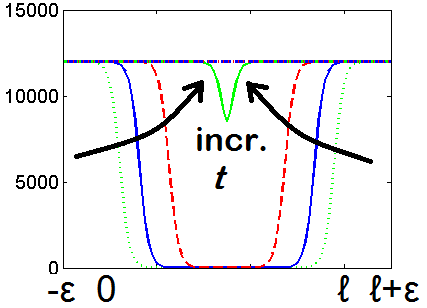}
        \includegraphics[width=1.5\linewidth]{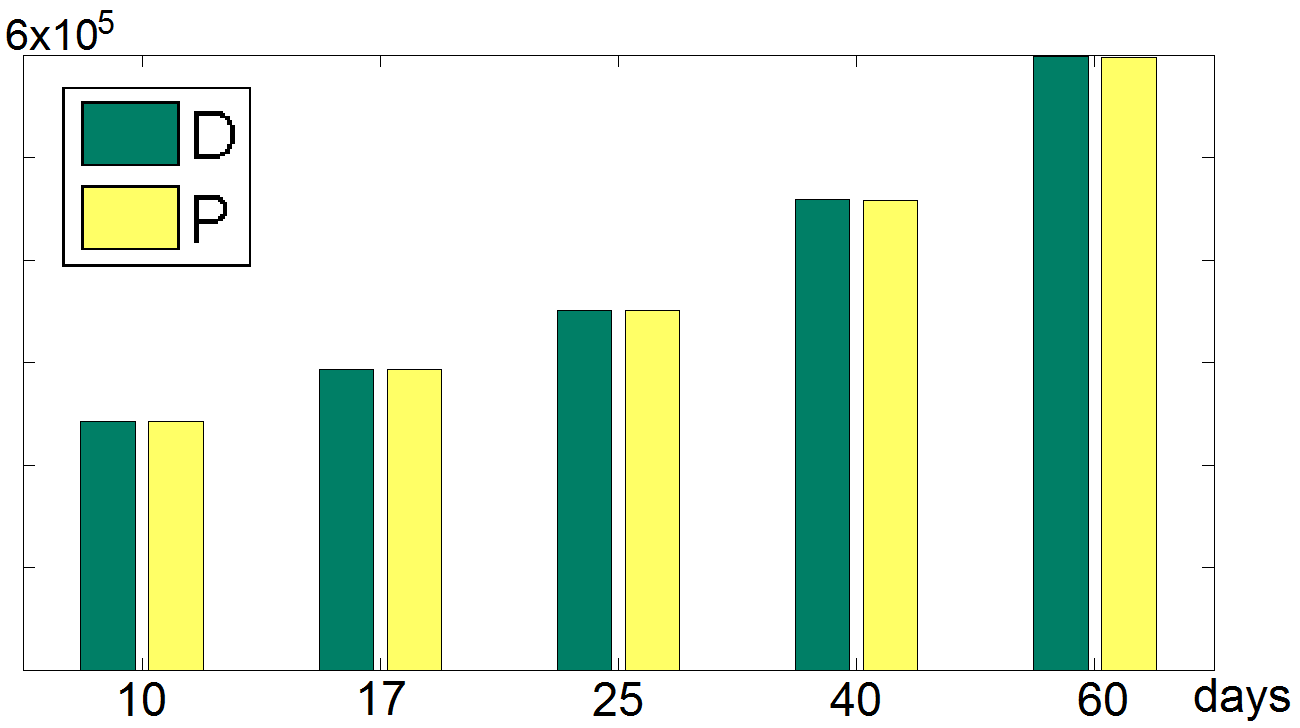}
        & \includegraphics[width=\linewidth]{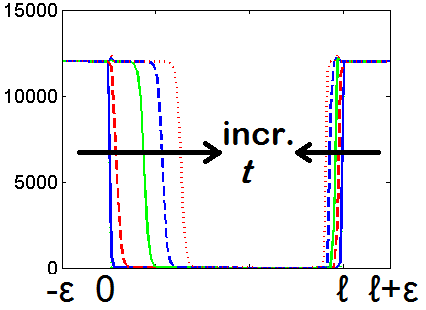}
        \includegraphics[width=1.5\linewidth]{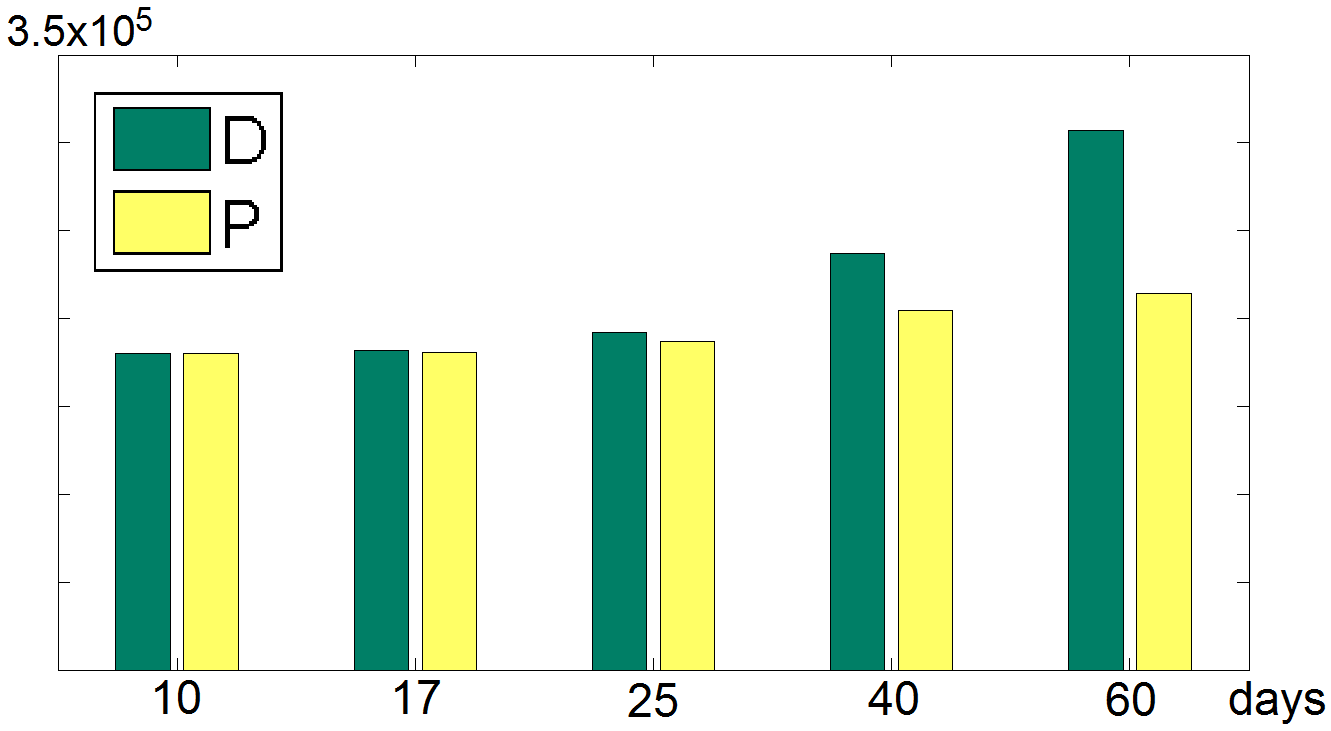}    \\
\hline
\raisebox{9\normalbaselineskip}[0pt][0pt]{\rot{\scalebox{3}{$\xi=1$}}}
        & \includegraphics[width=\linewidth]{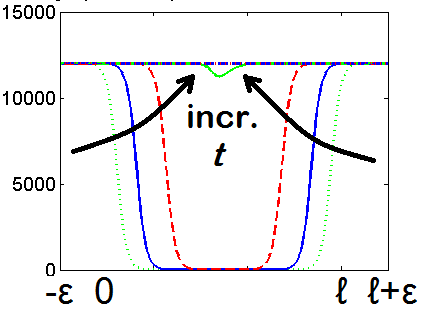}
        \includegraphics[width=1.5\linewidth]{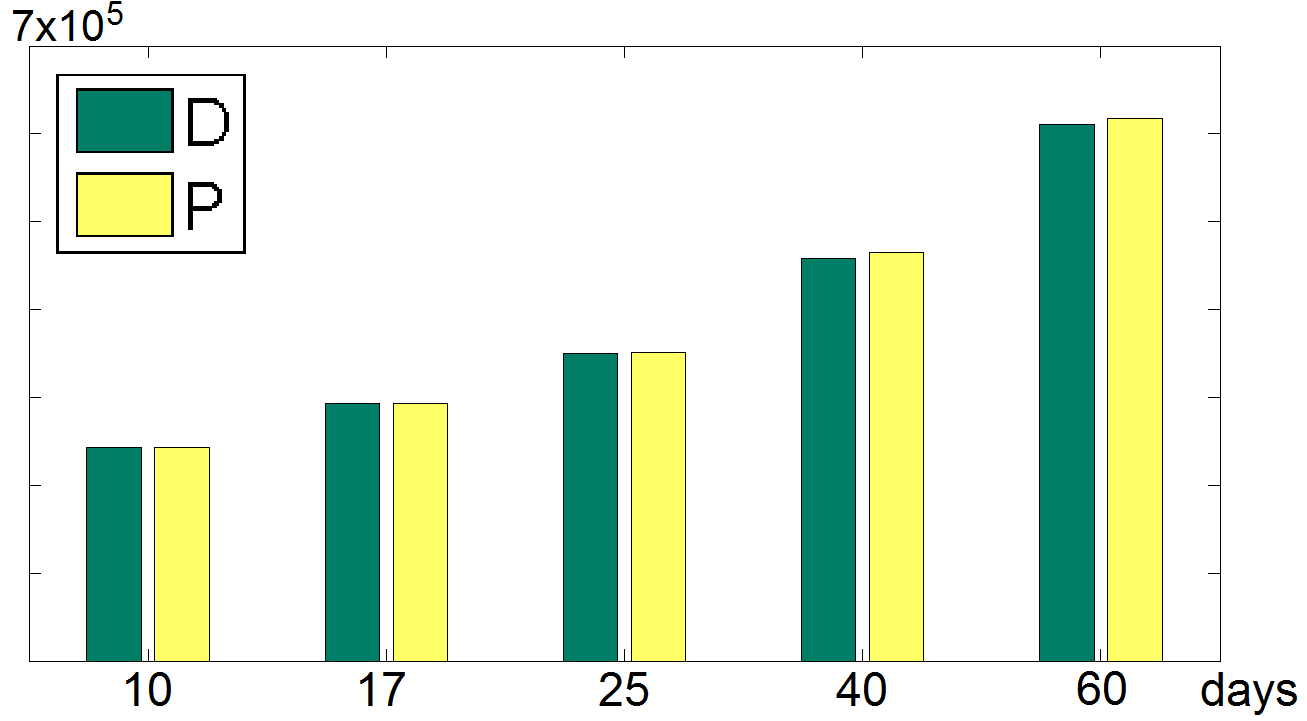}
        & \includegraphics[width=\linewidth]{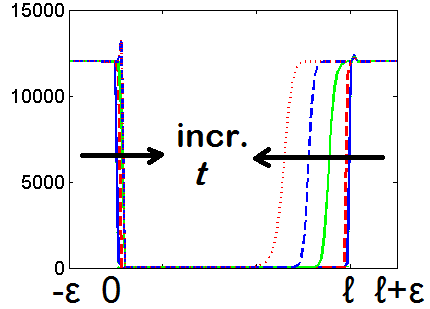}
        \includegraphics[width=1.5\linewidth]{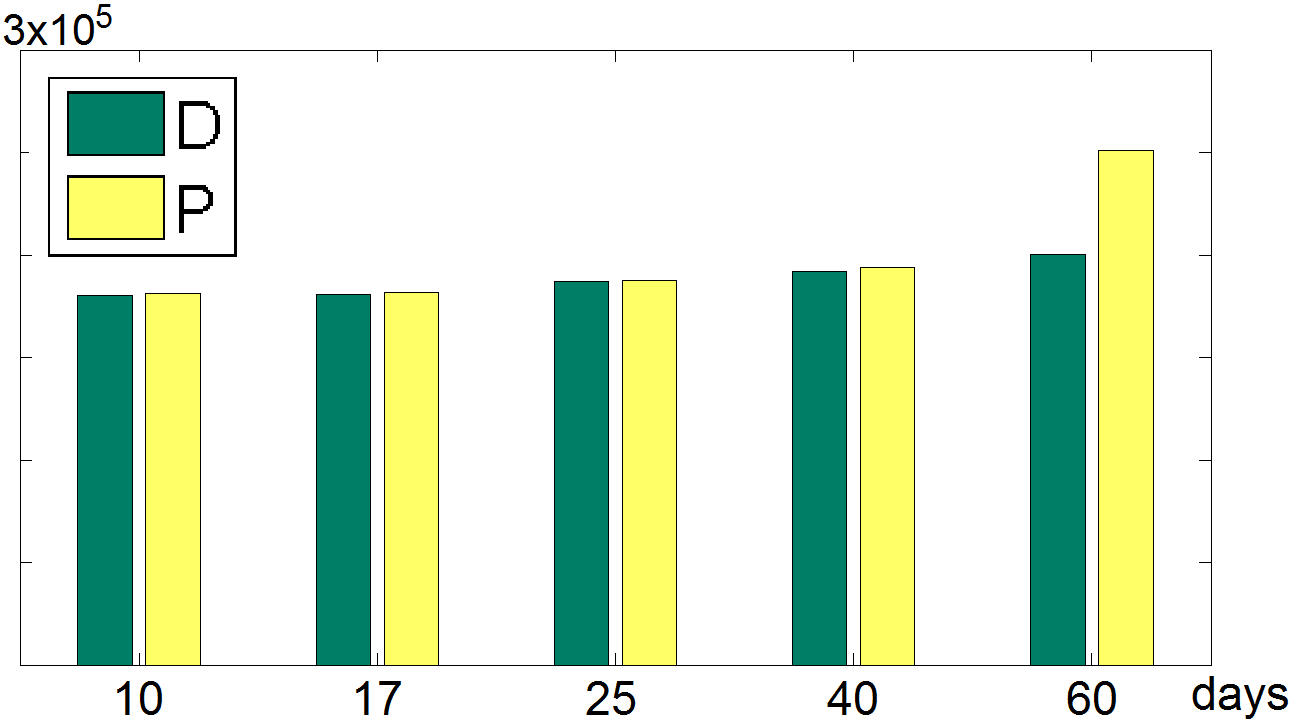}    \\
\hline
\end{tabular}}

\end{table}

In the self-organising case, we observe that varying $\xi$ between 0 and 0.75 does not significantly affect the model output for capillary regeneration; on the other hand, the initial conditions play a crucial role, since for a shallow wound ($b=5$) the lymphatic network is almost completely restored by day 60, while almost no healing is observed in the deep wound ($b=100$) scenario. In addition, lymphangiogenesis happens in a fairly symmetric fashion.
However, things appear to be quite different for $\xi=1$: in this case, both shallow and deep wounds exhibit a left-to-right lymphangiogenic process, which is completed by day 60. Note that while lymphangiogenesis occurs \emph{exclusively} from left to right in the deep wound scenario, in the shallow wound some lymphatic regeneration is also visible from the right-hand-side of the wound; this confirms our first observation that in a shallow wound logistic remodelling plays a more prominent role than in the deep wound setting.
These results suggest that the self-organising hypothesis is supported by the assumption that lymph flow, rather than interstitial flow, is the main contributor to advection in the wound space.

For the sprouting case, things are almost identical to the self-organising case for $0\leq \xi \leq 0.75$ and $b=5$ (shallow wound scenario). However, varying $\xi$ in this range seems to proportionally increase the left-to-right regeneration speed in the deep wound case ($b=100$), although it is still unable to account for complete regeneration at day 60. In addition, for $\xi=1$, while symmetric (although faster) healing is still visible for $b=5$, a capillary front advancing from right to left emerges in the deep wound scenario, though again this is not fast enough to restore the network by day 60.
This apparent ``switch'' of behaviour can be explained as follows.
$\xi=1$ corresponds to an advection component due exclusively to lymph flow coming from interrupted capillaries; hence, where $C_{op}=0$ both cells and chemicals tend to accumulate on one side of the wound. In the self-organising case, however, LECs display random movement and allow the capillary front to move. In the sprouting scenario, on the other hand, capillary tips are not subject to either diffusion or advection; therefore, the front of open capillaries tends to be stuck on the left-hand-side of the wound and chemotaxis tends to happen from right to left.
Thus, there is not such an obvious correlation between the value of $\xi$ and the validity of the sprouting hypothesis, in contrast to what we have seen above for the self-organising case. In the sprouting case, a very precise balance of lymph and interstitial flow is required to give a left-to-right lymphangiogenesis which is ``fast enough'', that is one which completes by day 60.

\begin{figure}[p]
\includegraphics[width=0.32\textwidth]{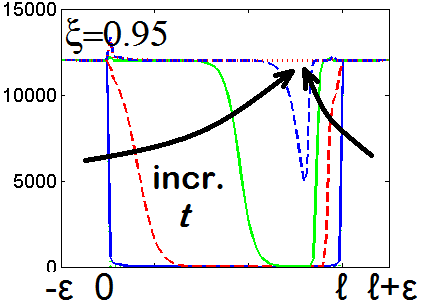}
\includegraphics[width=0.32\textwidth]{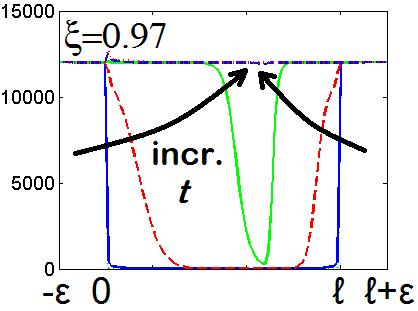}
\includegraphics[width=0.32\textwidth]{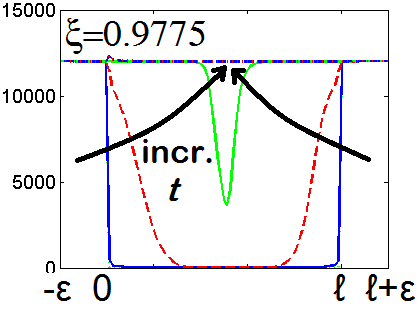}
\includegraphics[width=0.32\textwidth]{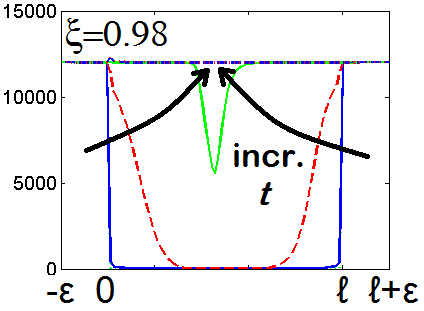}
\includegraphics[width=0.32\textwidth]{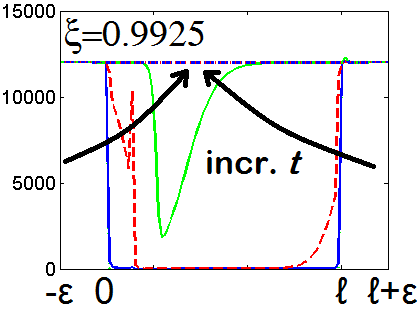}
\includegraphics[width=0.32\textwidth]{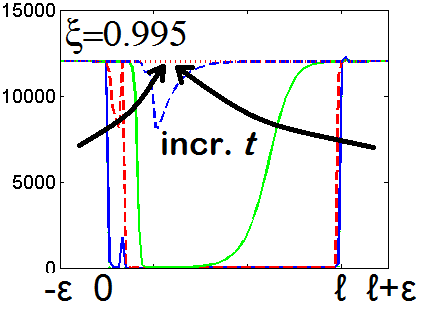}
\caption{Sprouting case: capillary density dynamics at different times for different values of $\xi$ in the range 0.75 -- 1 ($\xi=0.95,0.97,0.9775,0.98,0.9925,0.995$ -- from left to right, top to bottom, respectively), with $b=100$ (deep wound scenario). Arrows mark the direction of increasing $t$ in the simulations. }
\label{fig:sproutExtraXis}
\end{figure}

To further investigate the ``switch'' of behaviour (from left-to-right to vice-versa) observed in Table \ref{tab:cfrScapill} for $b=100$, we run some extra simulations of this case for $0.75<\xi<1$. Results are reported in Figure \ref{fig:sproutExtraXis} (note that no significant difference is observed for $0.75<\xi<0.95$, thus we report extra simulations only for values of $\xi$ starting from 0.95).
The simulations in Figure \ref{fig:sproutExtraXis} suggest that sprouting lymphangiogenesis switches from being left-to-right to being right-to-left as $\xi$ increases from 0 to 1, passing through symmetrical healing at around $\xi=0.9775$.

Therefore, the most ``realistic'' value for $\xi$ seems to be $\xi_O\approx 1$ for the self-organising case, and $\xi_S\approx 0.97$ for the sprouting case. For these values, 
the self-organising case predicts total healing by day 60 (in accordance with the data discussed in section \ref{sec:DataCmp}), while the sprouting case is a bit delayed in this respect. However, all the variables go back to their equilibrium levels in the latter case, while TGF-$\beta$, macrophages and VEGF stay at a high concentration in the right-hand-side of the wound in the self-organising scenario, which is not what we would expect  to happen in reality (simulations not shown).

\paragraph{Quantitative observations}

In order to make these observations more quantitative and compare them directly with the data sets presented in section \ref{sec:DataCmp}, in each case (i.e. both hypotheses and both combinations of $\xi$ and $b$ values) we calculate a parameter $\pi_{60}$ to quantify the percentage of healing/lymphatic regeneration at day 60. We also count how many LECs are present in the left (distal) and right (proximal) half of the domain at days 10, 17, 25, 40 and 60; in this way, we can directly compare the model output with the empirical data reported in Figure \ref{fig:LECdata}.

To define the quantity $\pi_{60}$, we consider one slice of the wound space, as depicted in Figure \ref{fig:domain}; we then consider the ratio between the space occupied by the lymphatic capillaries at day 60 and the original wound space. Thus, we consider
\begin{equation} \label{eq:def-pi60}
\pi_{60} = 100 \cdot \frac{S_{C,60}-S_{IC}}{S_{wound}} \; ,
\end{equation}
where $S_{C,60}$ is calculated as the area under the $C$-curve at $t=60$ (approximated as a polygon using the numerical results shown above) and $S_{wound}=C^{eq}\cdot(\ell+2\varepsilon)-S_{IC}$; $S_{IC}$ denotes the area subtended by the capillary initial profile curve defined in \eqref{eq:PDEinitcondsTMVC}, with $\nu = C$.
In this way, we estimate the portion of the \emph{real} initial wound (i.e. excluding the pre-existing capillary density) occupied by capillaries at day 60.
The values of $\pi_{60}$ for the various cases considered above are reported in Table \ref{table:pi60s}.

\begin{table}[p]

\caption{ Values of $\pi_{60}$ (defined in \eqref{eq:def-pi60}) for different values of $\xi$ and $b$ in the self-organising (O) and sprouting (S) cases. }

\label{table:pi60s}

\resizebox{0.75\textwidth}{!}{%
\begin{tabular}{c|ccccc|}
\hline
      &  \multicolumn{5}{c}{\textsc{shallow wound} ($b=5$)} \\
$\xi$ &  0  &  0.25  &  0.5  &  0.75  &  1       \\
\hline
  O   & 97.2\% & 97.2\% & 97.2\%  & 97.3\% & 101.7\% \\
\hline
  S   & 97.1\% & 97.1\% & 97.1\% & 97.1\% & 99.2\% \\
\hline
\hline
   & \multicolumn{5}{c}{\textsc{deep wound} ($b=100$)} \\
$\xi$ &  0  &  0.25  &  0.5  &  0.75  &  1 \\
\hline
 O   & 5.1\% & 5.1\%  & 5.2\% & 5.2\% & 104.1\% \\
  S   & 6.6\% & 8.3\% & 10.5\% & 17.5\% & 10.6\% \\
\hline
\end{tabular}
}
\end{table}

From Table \ref{table:pi60s}, we can see clearly how at day 60 the lymphatic vasculature will be restored to a level of 97\% or more for any value of $\xi$ in the shallow wound simulations in both the self-organising and sprouting case. For a deeper wound, however, the lymphatic capillary population is restored only up to about 5\% in the self-organising case and up to about 17\% in the sprouting case for $\xi\leq 0.75$; also, while the parameter $\pi_{60}$ has more or less the same value for all these $\xi$'s in the self-organising case, we observe an increase in $\pi_{60}$ for increasing $\xi$ in the sprouting scenario (from 6\% to 17\%). For $\xi=1$, though, the healing predictions are quite different: in the self-organising case, lymphatic capillary density slightly exceeds 100\% healing, while the sprouting case exhibits a capillary regeneration that covers only 10\% of the original wound.


To compare the model predictions with the data reported in Figure \ref{fig:LECdata}, we plot the number of LECs (considered as $L+C$ and $E+C$ in the self-organising and sprouting case respectively) in the left (distal) and right (proximal) half of the domain at days 10, 17, 25, 40 and 60. Such numbers are reported as bars in Tables \ref{tab:cfrOcapill} and \ref{tab:cfrScapill} (right-hand-side of each box), which correspond exactly to the cases plotted in Tables \ref{tab:cfrOcapill} and \ref{tab:cfrScapill} as simulations.

Comparing and contrasting the bar plots reported in Tables \ref{tab:cfrOcapill} and \ref{tab:cfrScapill} (right-hand-side of each box) with the data sets in Figure \ref{fig:LECdata}, we see that the row corresponding to $\xi=1$ is by far the best match for the self-organising case (the other values of $\xi$ giving almost no difference between the distal and proximal LEC density in any day after wounding).
For the sprouting case, it is natural to make a different distinction: lymphatic regeneration is always predicted to happen symmetrically in a shallow wound; in a deep wound, a slight distal-biased LEC density is observed appearing at days 40 and 60 for $\xi\leq 0.75$, while for $\xi=1$ the LEC density in the proximal half of the wound overtakes that in the distal half by day 60.

These observations confirm our first intuition: the self-organising case requires a value of $\xi$ close to one in order to observe a realistically fast left-to-right lymphangiogenesis, while the sprouting hypothesis needs a value of $\xi$ between 0.75 and 1 to produce similarly good results.
This difference could be explained by the different mechanisms regulating lymphangiogenesis in each case. In the self-organising hypothesis, capillaries form from LEC self-aggregation and disposition in capillary structures once these are (locally) sufficiently abundant; a constant ever-going interstitial flow slows this down because it prevents local LEC accumulation. In contrast lymph flow occurs only nearby interrupted capillary fronts, which move on as LECs coalesce into vessels. In the sprouting case, by contrast, the total absence of interstitial flow is a problem because neither capillary tips nor well-formed capillaries are subject to either random movement or lymph flow from interrupted capillaries; hence, interstitial flow is the only movement-inducing force, aside from chemical gradients. Moreover, in order to observe a chemical concentration peak on the right-hand-side of the wound (which, by chemotaxis, would induce a left-to-right migration of capillary tips), a good balance is required between an everywhere-present interstitial flow and a locally-active lymph flow. 


\subsection{No advection and additive advection cases} \label{sec:NoAdvAddAdv}

We now consider two final cases: that of no advection at all and that with  \emph{additive} advection (that is, where the advection velocity is as in \eqref{eq:def-lambdaChem} and \eqref{eq:def-lambdaCell}, but without the coefficients involving the parameter $\xi$).

Simulations of the no advection case are shown in Table \ref{tab:NOadv} for both  the self-organising and sprouting models.
\begin{table}[h]

\caption{Simulation of the self-organising (O) and sprouting (S) systems with parameters from Table \ref{table:param}  and initial condition as defined in \ref{sec:PDEICandBCs} where the advection terms are switched to zero. Arrows mark the direction of increasing $t$ in the simulations. }

\label{tab:NOadv}

\resizebox{0.8\textwidth}{!}{%
\begin{tabular}{ m{1cm} |c|c|}
\hline
 & \scalebox{2.5}{  \textsc{\Large shallow wound ($b=5$)} }  &  \scalebox{2.5}{ \textsc{\Large deep wound ($b=100$)} }  \\
\hline
\raisebox{10\normalbaselineskip}[0pt][0pt]{\scalebox{2}{\Huge O}}
        & \includegraphics[width=0.9\linewidth]{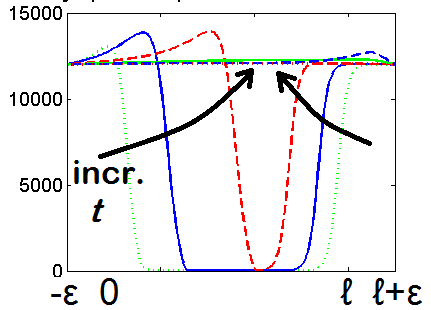}
        &  \includegraphics[width=0.9\linewidth]{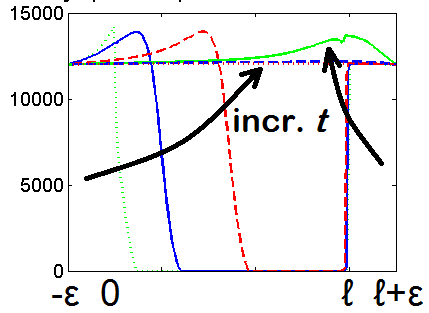}    \\
\hline
\raisebox{10\normalbaselineskip}[0pt][0pt]{\scalebox{2}{\Huge S}}
        & \includegraphics[width=0.9\linewidth]{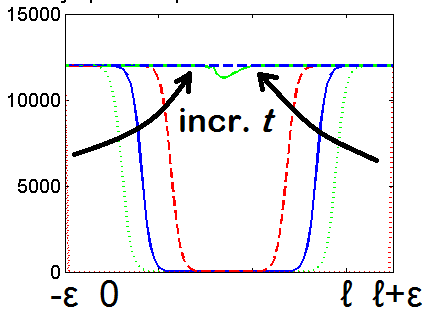}
        &  \includegraphics[width=0.9\linewidth]{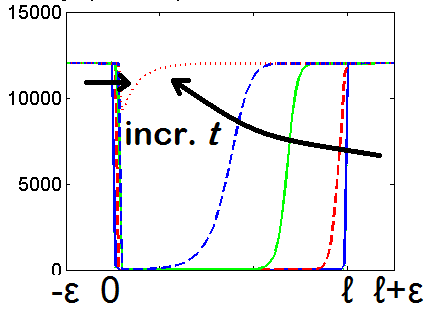}    \\
\hline
\end{tabular}
}

\end{table}
Note that, in the self-organising case, dynamics in the absence of advection resemble those reported in Table \ref{tab:cfrOcapill} for $\xi=1$, although here capillary regeneration is a bit slower.
Here the driving force behind left-to-right lymphangiogenesis is the influx of macrophages (which produce VEGF) and LECs (which form capillaries) from the left side of the interrupted capillaries. (Recall the influx term from the right edge is zero for macrophages and very small for LECs.)

In the sprouting case, too, capillary density evolution reflects that previously observed for $\xi=1$ (see Table \ref{tab:cfrScapill}). However, contrary to the self-organising case, here dynamics are significantly faster in the absence of advection. The reason behind this may lie in macrophages accumulating on the left side on the wound: consequently so does VEGF, which then drives the capillary sprouting from the right towards the peak on the left. Observe that here there are no LECs coming from the open capillaries on the left, so the regeneration is solely directed by gradients (capillary tips move towards increasing gradients of VEGF).

These results suggest two conclusions regarding advection:
\begin{enumerate}
\item advection contributes to the \emph{speed} of the lymphatic regeneration, speeding up the process in the self-organising case and slowing it down under the sprouting hypothesis;
\item advection is of greater importance in the sprouting case, where it actually determines the direction (left-to-right or vice versa) in which healing occurs.
\end{enumerate}
In other words, while the self-organising hypothesis seems to be able to explain left-to-right lymphangiogenesis on its own (thanks to the free LECs influx, primarily from the left side), the sprouting system needs some kind of force pushing VEGF towards the right of the domain so as to form a gradient driving capillary sprouts from the distal to the proximal end of the wound.


Finally, we investigate what happens when the advection velocities for chemicals and cells are replaced, respectively, by
\begin{eqnarray}
\lambda^{chem}(C_{op}) & = & ( \lambda_1^{chem} \cdot C_{op} ) + \lambda_2^{chem} \; \mbox{ and }
 \label{eq:def-lambdaChemADD}   \\
\lambda^{cell}(C_{op}) & = & ( \lambda_1^{cell} \cdot C_{op} ) + \lambda_2^{cell} \; ;
 \label{eq:def-lambdaCellADD}
\end{eqnarray}
we call this the \emph{additive advection} case. 
%
%
%
%
This time lymphangiogenesis, while not appearing overly affected in the shallow wound case, is heavily slowed down in the deep wound scenario (simulation not shown).
This reflects the fact that, when lymphatic regeneration is driven mainly by chemical gradients, a sufficiently strong advection force has a negative effect in healing because it does not allow chemicals and cells to accumulate and thereby produce sufficiently steep gradients.


\subsection{Overall comparison of O and S} \label{sec:OverallCmpr}

Here the overall similarities and differences between the self-organising and sprouting hypotheses are summarised:
\begin{itemize}

\item In shallow wounds lymphangiogenesis appears to be dominated by logistic growth/remodelling and occurs symmetrically from both sides of the wound.
In this case, there is little difference between the two hypotheses in terms of the dynamics of wound healing lymphangiogenesis.

\item Steeper initial conditions (as in a deep wound) lead to slower capillary regeneration; this is reasonable, since smaller/shallower wounds are expected to heal faster \cite{vowden2011,zimny2002} (see also \cite{monstrey2008} for burn depth).
In the deep wound case we also observe a marked difference in behaviour between the two hypotheses: the self-organising case exhibits a very slow progression for values of $\xi$ not close to 1, with the empirically observed speed occurring for $\xi=1$; by contrast, the sprouting hypothesis predicts lymphangiogenesis to take place from left to right at a speed that increases with $\xi$ up to $\xi\approx 0.9775$, when it becomes symmetric; for larger $\xi$ healing switches to a right-to-left process, at decreasing speed as $\xi$ approaches 1. This variety of behaviour highlights how important the ``balance'' between interstitial and lymph flow is in the advection terms \eqref{eq:def-lambdaChem} and \eqref{eq:def-lambdaCell}.

\end{itemize}


\section{Discussion}  

The results presented in this paper provide new insights in the understanding of lymphangiogenesis mechanisms.
Wound healing lymphangiogenesis is increasingly considered a fundamental aspect of the regeneration process, but there is still no consensus in the scientific community about how this phenomenon takes place. In particular, two main hypotheses have been advanced to describe the lymphangiogenesis process: the self-organising hypothesis \cite{benest2008,rutkowski2006} and the sprouting hypothesis \cite{norrmen2011,tammela2010}. Here we propose two different PDE systems to describe the two sets of  assumptions.
The present work shows how the problem of determining the exact lymphatic regeneration mechanism is intertwined with another open question in cellular biology: is interstitial flow a determining factor in cell migration? \cite{rutkowski2007} In this paper we explore the more general case of the effects of advection due to the combination of interstitial flow and lymph flow coming from the interrupted capillaries.
In addition, we also consider how different initial conditions, corresponding to shallow and deep wounds, affect the healing process.

The numerical simulations of the two systems we propose as describers for the self-organising and sprouting hypotheses suggest that the observation of left-to-right lymphangiogenesis does not justify \emph{per se} the self-organising hypothesis: our sprouting-hypothesis system can also reproduce this phenomenon, although for a very precise balance of lymph and interstitial flow.
Therefore, a reliable value of $\xi$ is needed  in order to choose between the two hypotheses. Other discriminating factors are that:
\begin{itemize}
\item capillary density in the sprouting case never significantly exceeds its normal value $C^{eq}$, while overcoming this value is predicted in the self-organising case;
\item in the self-organising case there is an excess of TGF-$\beta$, macrophages and VEGF persisting downstream of the lymph flow after capillaries have reached their healthy equilibrium level.
\end{itemize}

Biologically, it is not clear which is the main contributor to advection between interstitial flow and lymph flow coming from the interrupted capillaries; the models that we have presented suggest the latter is more relevant, and that the value of $\xi$ is above 0.75 in both modelled hypotheses. Moreover, our simulations hint at an inhibiting action of interstitial flow on lymphangiogenesis: strong interstitial flow here seems to significantly slow down capillary regeneration. This may be attributed to the fact that a ubiquitous advection force prevents chemical gradients from forming on the ``correct'' side of the wound.

Finally, initial conditions (that is, the type of wound, shallow or deep) strongly affect the speed and shape of the regeneration process: deeper wounds require more time to heal, and lymphangiogenesis will occur more markedly in the direction of the lymph flow in this case.

Our results emphasise the importance of advection in tissue regeneration; this concept could be of particular importance in describing the emerging concept of \emph{autologous chemotaxis}, that is the phenomenon whereby a cell can receive directional cues while at the same time being the source of such cues (see \cite{rutkowski2007,shields2007}).

Further developments of the model could include the blood vasculature, so to allow a direct comparison between the regenerations of the two vessel structures. The model could also be adapted to investigate differences in lymphatic regeneration in a diabetic scenario, as in \cite{bianchi2015}. It would also be interesting to investigate the similarities and the differences between wound healing lymphangiogenesis and tumour lymphangiogenesis: tumour cells are known to release lymphangiogenic factors and the tumour mass alters tissue pressure and interstitial flow, which could in turn promote pathological lymphangiogenesis in cancer \cite{cao2005b,christiansen2011,lunt2008,rofstad2014,simonsen2012}.

A definitive answer to the question of whether the self-organising or sprouting hypothesis better describes lymphangiogenesis will require a more informed evaluation of the relative contribution of interstitial and lymph flow to advection in the wound space, and more detailed spatio-temporal measures of capillary density and chemical concentrations: do we observe a ``bump'' exceeding normal capillary density along the capillary healing front? Do TGF-$\beta$, macrophages and VEGF persist at a high level downstream of the lymph flow after lymphatic regeneration is complete?

\section*{Acknowledgements}
A.B. was funded in part by a Maxwell Institute Scholarship from Heriot-Watt University.
K.J.P. acknowledges partial support from BBSRC grant BB/J015940/1.



\appendix

\section{Parameter estimation}  \label{appPARpde}

\subsection{Sizes, weights, equilibria and velocities} \label{app-sizes}

\subsubsection{Domain size}
We consider a full-thickness wound of length $\ell=5\mbox{ mm}$, inspired by \cite{zheng2007}. 
For the surrounding skin, we consider a (small) variable width $\varepsilon$. Thus, we have a domain of length $5 \mbox{ mm} + 2\varepsilon$.
In all the simulations reported in the present paper, $\varepsilon=1$; the nature of the observations does not change if a different value of $\varepsilon$ is chosen (simulations not shown).


\subsubsection{TGF-$\beta$ molecular weight and equilibrium $T^{eq}$} \label{app-TGFmwEquil}
We take TGF-$\beta$ molecular weight to be approximately 25 kDa \cite[active/mature isoform]{boulton1997,wakefield1988}.
The equilibrium value of active TGF-$\beta$ is about 30 pg/mm$^3$ \cite[Figure 2]{yang1999}.

\subsubsection{Macrophage volume and equilibrium $M^{eq}$}
A human alveolar macrophage has a volume $V_{M\Phi}$ of approximately $5000 \mu\mbox{m}^3 = 5\times 10^{-6} \mbox{mm}^3$ \cite{krombach1997}.
The macrophage steady state can be estimated from \cite[Figure 1]{weber1990},
which plots typical macrophage density in the skin.
This shows that there is an average of about 15 macrophages per 0.1mm$^2$ field.
Assuming a visual depth of 80 $\mu$m, the macrophage density becomes
15 cells/(0.1mm$^2\times 0.08$mm) = 1875 cells/mm$^3$.

\subsubsection{VEGF molecular weight and equilibrium $V^{eq}$}
VEGF molecular weight is taken to be 38 kDa \cite[VEGF-165]{kaur2012,yang2009}.
The VEGF equilibrium concentration is estimated to be 0.5 pg/mm$^3$ from \cite[Figure 1]{hormbrey2003} and \cite[Figure 2]{papaioannou2009}.

\subsubsection{Normal capillary density $C^{eq}$} \label{app-Cnorm}
In \cite{rutkowski2006} we find that ``it was not until \emph{day 60}, when functional and continuous lymphatic capillaries appeared normal'' and ``at \emph{day 60} the regenerated region had a complete lymphatic vasculature, the morphology of which appeared similar to that of native vessels''.
Hence, we assume that a capillary network that can be considered ``final'' appears at day 60, and we take $C^{eq}$ to be the number of LECs present at this time.
In \cite[Figure 2E]{rutkowski2006} we see that at that time there are about 80 cells.
This value corresponds to a 12 $\mu$m thin section.
In addition, from \cite[Figure 2D]{rutkowski2006} we can calculate the observed wound area, which is about $5.6\times 10^5 \, \mu\mbox{m}^2$.
In this way we get a volume of 0.0067 mm$^3$ with 80 cells, which corresponds to $C^{eq}=1.2\times 10^4$ cells/mm$^3$.

\subsubsection{Maximum capillary density $C_{max}$} \label{app-Cmax}

First of all, we want to convert 1 capillary section into a cell number. For this purpose, we assume EC cross-sectional dimensions to be those reported in \cite{haas1997}, namely $10\,\mu\mbox{m}\times 100\,\mu\mbox{m}$. We then assume that LECs lie ``longitudinally'' along the capillaries, and therefore only the short dimension contributes to cover or ``wrap'' the circumference of the capillary. Considering a capillary diameter of 55 $\mu$m as in \cite{fischer1996}, we have that each lymphatic capillary section is made of approximately 20 LECs (taking into account some overlapping).
Then, from \cite{vandenberg2003} we know that EC thickness is approximately 0.5 $\mu$m. Thus a capillary section is a circle of about $55+2\times 0.5 \, \mu$m diameter, corresponding, as described above, to 20 cells.

If we imagine stacking 1 mm$^3$ with capillaries of this size, we see that we can pile on $1\mbox{ mm}/ 56\,\mu\mbox{m} \approx 18$ layers of capillaries. Then, considering an EC length of 100 $\mu$m as in \cite{haas1997}, we have that 1 mm$^3$ fits at most a number of capillaries equivalent to the following amount of ECs:
$$
20 \mbox{ cells } \times 18 \times 18 \times \frac{1 \mbox{ mm}}{100 \, \mu\mbox{m}}
\approx 6.4 \times 10^4 \mbox{ cells } = C_{max} \; .
$$

\subsubsection{Lymph velocity}

\cite{fischer1996} suggests that the high lymph flow value (0.51mm/s) is due to high pressure following die injection. This suggests that a lower value (9.7 microns/s) might be considered as typical, in agreement with \cite{fischer1997}.
In both papers the normal lymph velocity seems to be around 10 microns/sec.
We thus assume lymph velocity to be $v_{lymph}$ = 10 micron/sec = 864 mm/day (from \cite{fischer1996,fischer1997}).




\subsubsection{Interstitial flow velocity}

First of all, we note that in \cite{rutkowski2007} interstitial flow in the skin is calculated to be around 10 microns/sec.
(Note that \cite{helm2005} is relevant for this aspect of our modelling, although it is less important for the estimation of parameters; in this reference the synergy between interstitial flow and VEGF gradient is discussed.)
Therefore, we will consider the interstitial flow to be also $v_{IF}$ = 10 microns/sec = 864 mm/day (from \cite{rutkowski2007}).


\subsection{Re-calculation of $s_M$ and $k_1$}

$s_M$ here is calculated in the same way as in \cite{bianchi2015}, but using our amended model equations presented here. 
For $k_1$, we point out that in \cite{bianchi2015} this parameter was appearing in the logistic part of the $M$-equation:
$ \nicefrac{d M}{dt} =  r_2 M - \nicefrac{r_2}{k_1}\cdot M^2$.
In the PDE systems we do not include such terms because only a minor fraction of macrophages undergo mitosis \cite{greenwood1973}.
However, death due to overcrowding is present in both models; comparing these terms, we see that our ``new'' $k_1$ corresponds to the ``old'' $k_1 / r_2$.

\subsection{Diffusion coefficients}

\subsubsection{VEGF diffusion coefficient $D_V$}
In \cite{miura2009} the authors observe that ``in general, the diffusion coefficient of protein molecules in liquid is of the order of $10^6\,\mu\mbox{m}^2/\mbox{h}=24\,\mbox{mm}^2/\mbox{day}$. This intuitively means that a molecule moves 10 $\mu$m/sec. To generate a gradient over the order of 100 $\mu$m, the timescale of protein decay should be around 10 seconds. In this specific case the protein decay time is about 1-10 hours. Therefore, the observed diffusion coefficient is too large and we need some mechanism to slow down the diffusion'' (where ``this specific case'' means that of VEGF).

In \cite{miura2009} the VEGF diffusion coefficient is estimated in three different ways: by a theoretical model ($0.24 \mbox{ mm}^2/\mbox{day}$), and by two different empirical techniques ($24 \mbox{ mm}^2/\mbox{day}$). The authors then suggest a diffusion coefficient of the order of $10^6\,\mu\mbox{m}^2/\mbox{h}=24 \mbox{ mm}^2/\mbox{day}$. However, they also used the same technique to determine the diffusion coefficient at the cell surface; this time the diffusion coefficient is estimated to be approximately $10^4\,\mu\mbox{m}^2/\mbox{h}=0.24\,\mbox{mm}^2/\mbox{day}$.
Keeping in mind all these considerations, for the model we take the intermediate value $D_V = 2.4\,\mbox{mm}^2/\mbox{day}$.

\subsubsection{TGF-$\beta$ diffusion coefficient $D_T$}
In \cite{lee2014} the authors estimate a TGF-$\beta$ diffusion coefficient of 0.36 mm$^2$/h = 8.64 mm$^2$/day  from \cite{brown1999,goodhill1997}.
In \cite{murphy2012} the authors estimate a TGF-$\beta$ diffusion coefficient of 2.54 mm$^2$/day using the Stokes-Einstein Formula.

We checked their consistency with the estimate for $D_V$ above.
The Stokes-Einstein equation of these calculated values assumes spherical particles of radius $r$ to have diffusion coefficient $D\sim{1}/{r}$; since the molecular weight $w$ of a particle is proportional to its volume, we have that $D\sim{1}/{\sqrt[3]{w}}$ and thus $D_T \approx 2.76$.

\subsubsection{Macrophage random motility $\mu_M$}
In\cite{farrell1990} we find ``Population random motility was characterized by the random motility coefficient, $\mu$, which was mathematically equivalent to a diffusion coefficient. $\mu$ varied little over a range of C5a [a protein] concentrations with a minimum of $0.86 \times 10^{-8} \mbox{cm}^2/\mbox{sec}$ in $1\times 10^{-7}$ M C5a to a maximum of $1.9\times 10^{-8} \mbox{cm}^2/\mbox{sec}$ in $1\times 10^{-11}$ M C5a''.
We thus take $\mu_M$ to be the average of these two values, that is $\mu_M = 1.38\times 10^{-8}\mbox{cm}^2/\mbox{s}\approx 0.12 \mbox{ mm}^2/\mbox{day}$.

\subsection{Advection parameters $\lambda_1$ and $\lambda_2$} \label{app-lambdas}

We will take $\lambda_2^{chem}$ to be equal to $v_{IF}$ calculated in \ref{app-sizes}; thus $\lambda_2^{chem}$ = 864 mm/day.
For $\lambda_1^{chem}$ it is more complicated, but we would say that if $C_{op}$ reaches the maximum possible value $C_{max}$ calculated in \ref{app-Cmax}, then $\lambda_1^{chem}\cdot C_{op} = v_{lymph}$, which was calculated in \ref{app-sizes}. That is, we assume that if the skin is ``packed'' with open capillaries, then the resulting flow will be the same as the usual lymph flow in the skin lymphatics). Hence $\lambda_1^{chem} = v_{lymph}/C_{max} = 0.0135 \mbox{ mm day}^{-1}\mbox{cell}^{-1}$.
For cells we assume smaller values due the higher friction that cells encounter in the tissue. In the absence of relevant empirical data, we take $\lambda_1^{cell}=\nicefrac{1}{10}\cdot\lambda_1^{chem}$ and $\lambda_2^{cell}=\nicefrac{1}{10}\cdot\lambda_2^{chem}$.

\subsection{Rate at which TGF-$\beta$ is internalised by macrophages $\gamma_1$}
At equilibrium, $C=C^{eq}$ and thus $p(C)=0$. Therefore, the equation for $T$ at equilibrium becomes
$$
a_M M^{eq} (T_L+r_1M^{eq}) - d_1 T^{eq} - \gamma_1 T^{eq} M^{eq} = 0 \; ,
$$
which leads to
$$
\gamma_1 = \frac{a_M M^{eq} (T_L+r_1M^{eq}) - d_1 T^{eq}}{T^{eq} M^{eq}} \approx 0.0042 \, \frac{\mbox{mm}^3}{\mbox{cells}\cdot\mbox{day}}\; .
$$


\subsection{Chemotaxis parameters}

\subsubsection{Macrophage chemotactic sensitivity towards TGF-$\beta$ $\chi_1$}
In \cite[Table 1]{lijeon2002} the chemotaxis coefficients of neutrophils for different gradients of interleukin-8 are listed (ranging from $0.6\times 10^{-7}$ to $12\times 10^{-7}$ mm$^2\cdot$mL$\cdot$ng$^{-1}\cdot$s$^{-1}$). We take the intermediate value $\chi_1 = 5\times 10^{-7}\mbox{mm}^2\mbox{mL}\mbox{ ng}^{-1}\mbox{s}^{-1}\approx 4\times 10^{-2}\mbox{mm}^2(\mbox{pg/mm}^3)^{-1}\mbox{day}^{-1}$.
To compare this value with one from another source, we consider \cite[Figure 8]{tranquillo1988}: although the chemotaxis coefficient is shown to depend on the attractant concentration, an average value is $\chi = 150 \mbox{ cm}^2\mbox{sec}^{-1}\mbox{M}^{-1}\approx 5.18\times 10^{-2}\mbox{mm}^2(\mbox{pg/mm}^3)^{-1}\mbox{day}^{-1}$ (using the TGF-$\beta$ molecular weight found in \ref{app-TGFmwEquil}). This result is encouraging because it is of the same order of magnitude as the previous estimate.

\subsubsection{LEC chemotactic sensitivity towards VEGF $\chi_2$}
In \cite{barkefors2008} a quantification is made of the effects of FGF2 and VEGF165 on HUVEC and HUAEC chemotaxis.
In \cite[Figure 6A]{barkefors2008} it is reported that the total distance migrated per HUVEC  in response to a 50 ng/mL gradient of VEGFA165 was about 150 $\mu$m. Considering that the analysed area of the cell migration chamber was 800 $\mu$m long and that the experiment lasted 200 minutes, we can estimate the endothelial cell velocity to be 150/200 = 0.75 $\mu$m/min = 1.08 mm/day and the VEGF gradient to be 50 ng/mL / 800 $\mu$m = 62.50 (pg/mm$^3$)/mm.
Now, the flux $\mathcal{J}$ in our equation is given by $\mathcal{J} = \chi_2 L \frac{\partial V}{\partial x}$; however, $\mathcal{J}$ can also be seen as the product of the mass density and the velocity of the flowing mass \cite{FluidMechanics}. Therefore, with $L$ being our mass density, we have
$$
\mbox{cell velocity} = \chi_2 \frac{\partial V}{\partial x}
$$
and then we can use the previous calculations to estimate
$$
\chi_2 = \frac{\mbox{cell velocity}}{\mbox{VEGF gradient}} = \frac{1.08 \mbox{mm/day}}{62.50 \mbox{(pg/mm}^3\mbox{)/mm}}
       = 0.0173 \, \frac{\mbox{mm}^2}{\mbox{day}} \frac{\mbox{mm}^3}{\mbox{pg}} \; .
$$
In order to have realistic cell movement dynamics, $\chi_2$ is taken to be 10 times bigger. This can be justified by the fact that the aforementioned data refer to HUVECs, and LECs might be faster than these cell types.  A more suitable dataset for this parameter would be very useful to better inform this estimate, but we are not aware of such data.
Also, chemical gradients created \emph{in vitro} are usually different between those observed \emph{in vivo} and they are known to highly affect cell velocity.

\subsubsection{Density-dependence of the macrophage chemotactic sensitivity $\omega$}

The cell density-dependence of the macrophage velocity is given by the factor $1/(1+\omega M)$.
This velocity is maximal when $M$ is close to zero and we assume that it is halved when $M$ reaches its carrying capacity $k_1^{old}$ (that is, the parameter $k_1$ in \cite{bianchi2015}).
We therefore take $\omega$ to be the inverse of the macrophage carrying capacity $k_1^{old}$.

\subsection{Macrophage inflow $\phi_1$} \label{app-phi1param}

We expect $\phi_1$ to be proportional to the lymph flow (estimated in \ref{app-sizes} as $v_{lymph} = 864 \mbox{ mm day}^{-1}$) and macrophage presence in the lymph.
In the same source \cite{fischer1996} that we used to estimate $v_{lymph}$, it is reported that the mean capillary diameter is 55 $\mu$m. Thus about $2.05\mbox{ mm}^3$ of lymph pass through a capillary bi-dimensional section in 1 day.

In \cite{cao2005} we find that a mouse leukocyte count in the blood is approximately 3 to $8\times 10^6$ cells/mL, and that of these about $2\times 10^6$ are macrophages coming from the lymph nodes; so we have a macrophage density of $2\times 10^3\mbox{ cells/mm}^3$ in the lymph. Therefore, each day about $2.05 \mbox{ mm}^3 \times 2\times 10^3\mbox{ cells/mm}^3 = 4.11\times 10^3$ macrophages pass in one capillary. Converting capillaries into cell density as was done in \ref{app-Cmax}, we have an influx equal to $\frac{4.11}{20}\times 10^3\mbox{day}^{-1}=0.205\times 10^3 \mbox{day}^{-1}$.
However, the macrophage density reported in \cite{cao2005} refers to blood; we assume that this quantity in lymph (especially during inflammation) will be about 10 times bigger. Therefore, we will take $\phi_1 = 2.05\times 10^3 \mbox{day}^{-1}$.


\bibliographystyle{amsedit}
\bibliography{mybib-bioPDE,mybib-mathPDE}

\providecommand{\bysame}{\leavevmode\hbox to3em{\hrulefill}\thinspace}
\providecommand{\MR}{\relax\ifhmode\unskip\space\fi MR }
\providecommand{\MRhref}[2]{%
  \href{http://www.ams.org/mathscinet-getitem?mr=#1}{#2}
}
\providecommand{\href}[2]{#2}
\begin{thebibliography}{100}

\bibitem{adams2007}
R.~H. Adams and K.~Alitalo, \emph{{{M}olecular regulation of angiogenesis and
  lymphangiogenesis}}, Nat. Rev. Mol. Cell Biol. \textbf{8} (2007), no.~6,
  464--478.

\bibitem{ambrose2006}
C.~T. Ambrose, \emph{{{I}mmunology's first priority dispute--an account of the
  17th-century {R}udbeck-{B}artholin feud}}, Cell. Immunol. \textbf{242}
  (2006), no.~1, 1--8.

\bibitem{asai2012}
J.~Asai, H.~Takenaka, S.~Hirakawa, J.~Sakabe, A.~Hagura, S.~Kishimoto,
  K.~Maruyama, K.~Kajiya, S.~Kinoshita, Y.~Tokura, and N.~Katoh, \emph{Topical
  simvastatin accelerates wound healing in diabetes by enhancing angiogenesis
  and lymphangiogenesis}, Am. J. of Pathol. \textbf{181} (2012), no.~6,
  2217--2224.

\bibitem{barkefors2008}
I.~Barkefors, S.~Le~Jan, L.~Jakobsson, E.~Hejll, G.~Carlson, H.~Johansson,
  J.~Jarvius, J.~W. Park, N.~Li~Jeon, and J.~Kreuger, \emph{{{E}ndothelial cell
  migration in stable gradients of vascular endothelial growth factor {A} and
  fibroblast growth factor 2: effects on chemotaxis and chemokinesis}}, J.
  Biol. Chem. \textbf{283} (2008), no.~20, 13905--13912.

\bibitem{benest2008}
A.~V. Benest, S.~J. Harper, S.~Y. Herttuala, K.~Alitalo, and D.~O. Bates,
  \emph{{{V}{E}{G}{F}-{C} induced angiogenesis preferentially occurs at a
  distance from lymphangiogenesis}}, Cardiovasc. Res. \textbf{78} (2008),
  no.~2, 315--323.

\bibitem{bernatchez1999}
P.~N. Bernatchez, S.~Soker, and M.~G. Sirois, \emph{Vascular endothelial growth
  factor effect on endothelial cell proliferation, migration, and
  platelet-activating factor synthesis is {F}lk-1-dependent}, J. Biol. Chem.
  \textbf{274} (1999), no.~43, 31047--31054.

\bibitem{bianchi2015}
A.~Bianchi, K.~J. Painter, and J.~A. Sherratt, \emph{A mathematical model for
  lymphangiogenesis in normal and diabetic wounds}, Journal of Theoretical
  Biology \textbf{383} (2015), 61--86.

\bibitem{boardman2003}
K.~C. Boardman and M.~A. Swartz, \emph{Interstitial flow as a guide for
  lymphangiogenesis}, Circ. Res. \textbf{92} (2003), 801--808.

\bibitem{boulton1997}
R.~Boulton, A.~Woodman, D.~Calnan, C.~Selden, F.~Tam, and H.~Hodgson,
  \emph{{{N}onparenchymal cells from regenerating rat liver generate
  interleukin-1alpha and -1beta: a mechanism of negative regulation of
  hepatocyte proliferation}}, Hepatology \textbf{26} (1997), no.~1, 49--58.

\bibitem{brem2007}
H.~Brem and M.~Tomic-Canic, \emph{{{C}ellular and molecular basis of wound
  healing in diabetes}}, J. Clin. Invest. \textbf{117} (2007), no.~5,
  1219--1222.

\bibitem{brown1999}
D.~R. Brown, \emph{Dependence of neurones on astrocytes in a coculture system
  renders neurones sensitive to transforming growth factor {$\beta$}1-induced
  glutamate toxicity}, Journal of Neurochemistry \textbf{72} (1999), no.~3,
  943--953.

\bibitem{byrne1995}
H.~M. Byrne and M.~A.~J. Chaplain, \emph{Explicit solutions of a simplified
  model of capillary sprout growth during tumour angiogenesis}, Appl. Math.
  Lett. \textbf{8} (1995), no.~5, 71--76.

\bibitem{byrne2000}
H.~M. Byrne, M.~A.~J. Chaplain, D.~L. Evans, and I.~Hopkinson,
  \emph{Mathematical modelling of angiogenesis in wound healing: Comparison of
  theory and experiment}, J. Theor. Med. \textbf{2} (2000), no.~3, 175--197.

\bibitem{cao2005}
C.~Cao, D.~A. Lawrence, D.~K. Strickland, and L.~Zhang, \emph{{{A} specific
  role of integrin {M}ac-1 in accelerated macrophage efflux to the
  lymphatics}}, Blood \textbf{106} (2005), no.~9, 3234--3241.

\bibitem{cao2005b}
Y.~Cao, \emph{{{O}pinion: emerging mechanisms of tumour lymphangiogenesis and
  lymphatic metastasis}}, Nat. Rev. Cancer \textbf{5} (2005), no.~9, 735--743.

\bibitem{castiglioni1947}
A.~Castiglioni, \emph{A history of medicine}, Alfred A. Knopf, New York, 1947.

\bibitem{cho2006}
C.-H. Cho, H.-K. Sung, K.-T. Kim, H.~G. Cheon, G.~T. Oh, H.~J. Hong, O.-J. Yoo,
  and G.~Y. Koh, \emph{{C}{O}{M}{P}-angiopoietin-1 promotes wound healing
  through enhanced angiogenesis, lymphangiogenesis, and blood flow in a
  diabetic mouse model}, PNAS \textbf{103} (2006), no.~13, 4946--4951.

\bibitem{choi2012}
I.~Choi, S.~Lee, and Y.~K. Hong, \emph{{{T}he new era of the lymphatic system:
  no longer secondary to the blood vascular system}}, Cold Spring Harb.
  Perspect. Med. \textbf{2} (2012), no.~4, a006445.

\bibitem{christiansen2011}
A.~Christiansen and M.~Detmar, \emph{{{L}ymphangiogenesis and cancer}}, Genes
  Cancer \textbf{2} (2011), no.~12, 1146--1158.

\bibitem{cobbold2000}
C.~A. Cobbold and J.~A. Sherratt, \emph{Mathematical modelling of nitric oxide
  activity in wound healing can explain keloid and hypertrophic scarring}, J.
  Theor. Biol. \textbf{204} (2000), 257--288.

\bibitem{decrescenzo2001}
G.~De~Crescenzo, S.~Grothe, J.~Zwaagstra, M.~Tsang, and M.~D. O'Connor-McCourt,
  \emph{Real-time monitoring of the interactions of transforming growth
  factor{-$\beta$} {(TGF-$\beta$)} isoforms with latency-associated protein and
  the ectodomains of the {TGF-$\beta$} type {II} and {III} receptors reveals
  different kinetic models and stoichiometries of binding}, J. Biol. Chem.
  \textbf{276} (2001), no.~32, 29632--29643.

\bibitem{dodson1924}
J.~F. Dodson, \emph{Herophilus of alexandria}, Proc. R. Soc. Med. \textbf{18}
  (1924-5), 19--32.

\bibitem{FluidMechanics}
J.~F. Douglas, J.~M. Gasiorek, J.~A. Swaffield, and L.~B. Jack, \emph{Fluid
  mechanics}, 5th ed., Prentice Hall, 2005.

\bibitem{drew2007}
P.~Drew, J.~Posnett, and L.~Rusling, \emph{{{T}he cost of wound care for a
  local population in {E}ngland}}, Intl. Wound. J. \textbf{4} (2007), no.~2,
  149--155.

\bibitem{edelstein1982}
L.~Edelstein, \emph{{T}he {P}ropagation of {F}ungal {C}olonies: {A} {M}odel for
  {T}issue {G}rowth}, J. Theor. Biol. \textbf{98} (1982), 679--701.

\bibitem{farrell1990}
B.~E. Farrell, R.~P. Daniele, and D.~A. Lauffenburger, \emph{{{Q}uantitative
  relationships between single-cell and cell-population model parameters for
  chemosensory migration responses of alveolar macrophages to {C}5a}}, Cell
  Motil. Cytoskeleton \textbf{16} (1990), no.~4, 279--293.

\bibitem{fischer1997}
M.~Fischer, U.~Costanzo, U.~Hoffmann, A.~Bollinger, and U.~K. Franzeck,
  \emph{{{F}low velocity of cutaneous lymphatic capillaries in patients with
  primary lymphedema}}, Int J Microcirc Clin Exp \textbf{17} (1997), no.~3,
  143--149.

\bibitem{fischer1996}
M.~Fischer, U.~K. Franzeck, I.~Herrig, U.~Costanzo, S.~Wen, M.~Schiesser,
  U.~Hoffmann, and A.~Bollinger, \emph{{{F}low velocity of single lymphatic
  capillaries in human skin}}, Am. J. Physiol. \textbf{270} (1996), no.~1~Pt~2,
  H358--363.

\bibitem{flegg2015}
J.~A. Flegg, S.~N. Menon, P.~K. Maini, and D.~L.~S. McElwain, \emph{On the
  mathematical modeling of wound healing angiogenesis in skin as a
  reaction-transport process}, Frontiers in Physiology \textbf{6} (2015),
  no.~262.

\bibitem{flegg2012}
J.~A. Flegg, H.~M. Byrne, M.~B. Flegg, and D.~L.~S. McElwain, \emph{Wound
  healing angiogenesis: the clinical implications of a simple mathematical
  model}, J. Theor. Biol. \textbf{300} (2012), 309--316.

\bibitem{fleury2006}
M.~E. Fleury, K.~C. Boardman, and M.~A. Swartz, \emph{Autologous morphogen
  gradients by subtle interstitial flow and matrix interactions}, Biophysical
  Journal \textbf{91} (2006), 113--121.

\bibitem{friedman2005}
A.~Friedman and G.~Lolas, \emph{Analysis of a mathematical model of tumor
  lymphangiogenesis}, Math. Models Methods Appl. Sci. \textbf{15} (2005),
  no.~1, 95--107.

\bibitem{galie2009}
P.~Galie and R.~L. Spilker, \emph{{{A} two-dimensional computational model of
  lymph transport across primary lymphatic valves}}, J Biomech Eng \textbf{131}
  (2009), no.~11, 111004.

\bibitem{goldman2007}
J.~Goldman, J.~M. Rutkowski, J.~D. Shields, M.~C. Pasquier, Y.~Cui, H.~G.
  Schmokel, S.~Willey, D.~J. Hicklin, B.~Pytowski, and M.~A. Swartz,
  \emph{{{C}ooperative and redundant roles of {V}{E}{G}{F}{R}-2 and
  {V}{E}{G}{F}{R}-3 signaling in adult lymphangiogenesis}}, FASEB J.
  \textbf{21} (2007), no.~4, 1003--1012.

\bibitem{goodhill1997}
G.~J. Goodhill, \emph{{D}iffusion in axon guidance}, Eur. J. Neurosci.
  \textbf{9} (1997), no.~7, 1414--1421.

\bibitem{gosiewska1999}
A.~Gosiewska, C.~Yi, O.~Blanc-Brude, and J.~C. Geesin, \emph{Characterization
  of a macrophage-based system for studying the activation of latent
  {TGF-$\beta$}}, Meth. Cell Sci. \textbf{21} (1999), 47--56.

\bibitem{grainger1995}
D.~J. Grainger, L.~Wakefield, H.~W. Bethell, R.~W. Farndale, and J.~C.
  Metcalfe, \emph{{R}elease and activation of platelet latent {T}{G}{F}-beta in
  blood clots during dissolution with plasmin}, Nat. Med. \textbf{1} (1995),
  no.~9, 932--937.

\bibitem{greenwood1973}
B.~Greenwood, \emph{The mitosis of sheep blood monocytes in tissue culture},
  Quart. J. Exp. Physiol. \textbf{58} (1973), 369--377.

\bibitem{haas1997}
T.~L. Haas and B.~R. Duling, \emph{Morphology favors an endothelial cell
  pathway for longitudinal conduction within arterioles}, Microvasc. Res.
  \textbf{53} (1997), 113--120.

\bibitem{helm2005}
C.~L. Helm, M.~E. Fleury, A.~H. Zisch, F.~Boschetti, and M.~A. Swartz,
  \emph{{{S}ynergy between interstitial flow and {V}{E}{G}{F} directs capillary
  morphogenesis in vitro through a gradient amplification mechanism}}, Proc.
  Natl. Acad. Sci. U.S.A. \textbf{102} (2005), no.~44, 15779--15784.

\bibitem{heppell2015}
C.~Heppell, T.~Roose, and G.~Richardson, \emph{A model for interstitial
  drainage through a sliding lymphatic valve}, Bull Math Biol \textbf{77}
  (2015), 1101--1131.

\bibitem{hormbrey2003}
E.~Hormbrey, C.~Han, A.~Roberts, D.~A. McGrouther, and A.~L. Harris, \emph{The
  relationship of human wound vascular endothelial growth factor {(VEGF)} after
  breast cancer surgery to circulating vegf and angiogenesis}, Clin. Cancer
  Res. \textbf{9} (2003), 4332--4339.

\bibitem{huggenberger2011}
R.~Huggenberger, S.~S. Siddiqui, D.~Brander, S.~Ullmann, K.~Zimmermann,
  M.~Antsiferova, S.~Werner, K.~Alitalo, and M.~Detmar, \emph{{{A}n important
  role of lymphatic vessel activation in limiting acute inflammation}}, Blood
  \textbf{117} (2011), no.~17, 4667--4678.

\bibitem{hyytiainen2004}
M.~Hyyti{\"a}inen, C.~Penttinen, and J.~Keski-Oja, \emph{{{L}atent
  {T}{G}{F}-beta binding proteins: extracellular matrix association and roles
  in {T}{G}{F}-beta activation}}, Crit Rev Clin Lab Sci \textbf{41} (2004),
  no.~3, 233--264.

\bibitem{jeffcoate2003}
W.~J. Jeffcoate and K.~G. Harding, \emph{{{D}iabetic foot ulcers}}, Lancet
  \textbf{361} (2003), no.~9368, 1545--1551.

\bibitem{ji2005}
R.~C. Ji, \emph{Characteristics of lymphatic endothelial cells in physiological
  and pathological conditions}, Histol. Histopathol. \textbf{20} (2005),
  155--175.

\bibitem{kaminska2005}
B.~Kaminska, A.~Wesolowska, and M.~Danilkiewicz, \emph{{TGF} beta signalling
  and its role in tumor pathogenesis}, Acta Biochimica Polonica \textbf{52}
  (2005), no.~2, 329--337.

\bibitem{kaur2012}
H.~Kaur and L.~Y. Yung, \emph{{{P}robing high affinity sequences of {D}{N}{A}
  aptamer against {V}{E}{G}{F}165}}, PLoS ONE \textbf{7} (2012), no.~2, e31196.

\bibitem{khalil1996}
N.~Khalil, S.~Corne, C.~Whitman, and H.~Yacyshyn, \emph{{{P}lasmin regulates
  the activation of cell-associated latent {T}{G}{F}-beta 1 secreted by rat
  alveolar macrophages after in vivo bleomycin injury}}, Am. J. Respir. Cell
  Mol. Biol. \textbf{15} (1996), no.~2, 252--259.

\bibitem{khalil1993}
N.~Khalil, C.~Whitman, L.~Zuo, D.~Danielpour, and A.~Greenberg,
  \emph{Regulation of alveolar macrophage transforming growth factor{-$\beta$}
  secretion by corticosteroids in bleomycin-induced pulmonary inflammation in
  the rat}, J. Clin. Invest. \textbf{92} (1993), 1812--1818.

\bibitem{kiriakidis2003}
S.~Kiriakidis, E.~Andreakos, C.~Monaco, B.~Foxwell, M.~Feldmann, and
  E.~Paleolog, \emph{{VEGF} expression in human macrophages is
  {NF-$\kappa$B-}dependent: studies using adenoviruses expressing the
  endogenous {NF-$\kappa$B} inhibitor {I$\kappa$B$\alpha$} and a
  kinase-defective form of the {I$\kappa$B} kinase 2}, J. Cell Sci.
  \textbf{116} (2003), no.~4, 665--674.

\bibitem{kleinheinz2010}
J.~Kleinheinz, S.~Jung, K.~Wermker, C.~Fischer, and U.~Joos, \emph{Release
  kinetics of {V}{E}{G}{F}$_{165}$ from a collagen matrix and structural matrix
  changes in a circulation model}, Head Face Med. (2010), 6--17.

\bibitem{krombach1997}
F.~Krombach, S.~Munzing, A.~M. Allmeling, J.~T. Gerlach, J.~Behr, and
  M.~Dorger, \emph{{{C}ell size of alveolar macrophages: an interspecies
  comparison}}, Environ. Health Perspect. \textbf{105 Suppl 5} (1997),
  1261--1263.

\bibitem{lee2014}
S.~Lee, H.~J. Hwang, and Y.~Kim, \emph{{{M}odeling the role of
  {T}{G}{F}-$\beta$ in regulation of the {T}h17 phenotype in the
  {L}{P}{S}-driven immune system}}, Bull. Math. Biol. \textbf{76} (2014),
  no.~5, 1045--1080.

\bibitem{levine2001}
H.~A. Levine, S.~Pamuk, B.~D. Sleeman, and M.~Nilsen-Hamilton,
  \emph{Mathematical modeling of capillary formation and development in tumor
  angiogenesis: Penetration into the stroma}, Bull. Math. Biol. \textbf{63}
  (2001), no.~5, 801--863.

\bibitem{lijeon2002}
N.~Li~Jeon, H.~Baskaran, S.~K. Dertinger, G.~M. Whitesides, L.~Van~de Water,
  and M.~Toner, \emph{{{N}eutrophil chemotaxis in linear and complex gradients
  of interleukin-8 formed in a microfabricated device}}, Nat. Biotechnol.
  \textbf{20} (2002), no.~8, 826--830.

\bibitem{lohela2009}
M.~Lohela, M.~Bry, T.~Tammela, and K.~Alitalo, \emph{{{V}{E}{G}{F}s and
  receptors involved in angiogenesis versus lymphangiogenesis}}, Curr. Opin.
  Cell Biol. \textbf{21} (2009), no.~2, 154--165.

\bibitem{louveau2015}
A.~Louveau, I.~Smirnov, T.~J. Keyes, J.~D. Eccles, S.~J. Rouhani, J.~D. Peske,
  N.~C. Derecki, D.~Castle, J.~W. Mandell, K.~S. Lee, T.~H. Harris, and
  J.~Kipnis, \emph{{{S}tructural and functional features of central nervous
  system lymphatic vessels}}, Nature \textbf{523} (2015), no.~7560, 337--341.

\bibitem{ludwig1979}
D.~Ludwig, D.~G. Aronson, and H.~F. Weinberger, \emph{{S}patial {P}atterning of
  the {S}pruce {B}udworm}, J. Math. Biology \textbf{8} (1979), 217--258.

\bibitem{lunt2008}
S.~J. Lunt, T.~M. Kalliomaki, A.~Brown, V.~X. Yang, M.~Milosevic, and R.~P.
  Hill, \emph{{{I}nterstitial fluid pressure, vascularity and metastasis in
  ectopic, orthotopic and spontaneous tumours}}, BMC Cancer \textbf{8} (2008),
  2.

\bibitem{gabhann2004}
F.~Mac~Gabhann and A.~S. Popel, \emph{Model of competitive binding of vascular
  endothelial growth factor and placental growth factor to {VEGF} receptors on
  endothelial cells}, Am. J. Physiol. Heart Circ. Physiol. \textbf{286} (2004),
  H153--H164.

\bibitem{macdonald2008}
A.~J. Macdonald, K.~P. Arkill, G.~R. Tabor, N.~G. McHale, and C.~P. Winlove,
  \emph{{{M}odeling flow in collecting lymphatic vessels: one-dimensional flow
  through a series of contractile elements}}, Am. J. Physiol. Heart Circ.
  Physiol. \textbf{295} (2008), no.~1, H305--313.

\bibitem{mantovani2004}
A.~Mantovani, A.~Sica, S.~Sozzani, P.~Allavena, A.~Vecchi, and M.~Locati,
  \emph{{{T}he chemokine system in diverse forms of macrophage activation and
  polarization}}, Trends Immunol. \textbf{25} (2004), no.~12, 677--686.

\bibitem{mantzaris2004}
N.~V. Mantzaris, S.~Webb, and H.~G. Othmer, \emph{Mathematical modeling of
  tumor-induced angiogenesis}, J. Math. Biol. \textbf{49} (2004), 111--187.

\bibitem{margaris2012}
K.~N. Margaris and R.~A. Black, \emph{{{M}odelling the lymphatic system:
  challenges and opportunities}}, J. R. Soc. Interf. \textbf{9} (2012), no.~69,
  601--612.

\bibitem{maruyama2007}
K.~Maruyama, J.~Asai, M.~Ii, T.~Thorne, D.~W. Losordo, and P.~A. D'Amore,
  \emph{Decreased macrophage number and activation lead to reduced lymphatic
  vessel formation and contribute to impaired diabetic wound healing}, Am. J.
  Pathol. \textbf{70} (2007), 1178--1191.

\bibitem{may1968}
M.~T.~G.~T. May, \emph{Galen on the usefulness of the parts of the body, part
  ii}, Cornell University Press, Ithaca, NY, 1968.

\bibitem{mendoza2003}
E.~Mendoza and G.~W. Schmid-Sch{\"o}nbein, \emph{A model for mechanics of
  primary lymphatic valves}, J. Biomech. Eng. \textbf{125} (2003), 407--414.

\bibitem{miura2009}
T.~Miura and R.~Tanaka, \emph{\emph{In vitro} vasculogenesis models revisited -
  measurement of {VEGF} diffusion in matrigel}, Math. Model. Nat. Phenom.
  \textbf{4} (2009), no.~4, 118--130.

\bibitem{monstrey2008}
S.~Monstrey, H.~Hoeksema, J.~Verbelen, A.~Pirayesh, and P.~Blondeel,
  \emph{Assessment of burn depth and burn wound healing potential}, Burns
  \textbf{34} (2008), 761--769.

\bibitem{muller1987}
G.~M{\"u}ller, J.~Behrens, U.~Nussbaumer, P.~B{\"o}hlen, and W.~Birchmeier,
  \emph{Inhibitory action of transforming growth factor {$\beta$} on
  endothelial cells}, PNAS \textbf{84} (1987), 5600--5604.

\bibitem{murphy2012}
K.~E. Murphy, C.~L. Hall, P.~K. Maini, S.~W. McCue, and D.~L.~S. McElwain,
  \emph{A fibrocontractive mechanochemical model of dermal wound closure
  incorporating realistic growth factor kinetics}, Bull. Math. Biol.
  \textbf{74} (2012), no.~5, 1143--1170. \MR{2909123}

\bibitem{nguyen2007}
V.~P. K.~H. Nguyen, S.~H. Chen, J.~Trinh, H.~Kim, B.~L. Coomber, and D.~J.
  Dumont, \emph{Differential response of lymphatic, venous and arterial
  endothelial cells to angiopoietin-1 and angiopoietin-2}, BMC Cell Biol.
  (2007), 8:10.

\bibitem{norrmen2011}
C.~Norrmen, T.~Tammela, T.~V. Petrova, and K.~Alitalo, \emph{{{B}iological
  basis of therapeutic lymphangiogenesis}}, Circulation \textbf{123} (2011),
  no.~12, 1335--1351.

\bibitem{nunes1995}
I.~Nunes, R.~L. Shapiro, and D.~B. Rifkin, \emph{Characterization of latent
  {TGF}-$\beta$ activation by murine peritoneal macrophages}, J. Immunol.
  \textbf{155} (1995), 1450--1459.

\bibitem{oi2004}
M.~Oi, T.~Yamamoto, and K.~Nishioka, \emph{Increased expression of
  {TGF}-$\beta$1 in the sclerotic skin in bleomycin-`susceptible' mouse
  strains}, J. Med. Dent. Sci. \textbf{51} (2004), 7--17.

\bibitem{oliver2002}
G.~Oliver and M.~Detmar, \emph{{{T}he rediscovery of the lymphatic system: old
  and new insights into the development and biological function of the
  lymphatic vasculature}}, Genes Dev. \textbf{16} (2002), no.~7, 773--783.

\bibitem{papaioannou2009}
A.~I. Papaioannou, E.~Zakynthinos, K.~Kostikas, T.~Kiropoulos, A.~Koutsokera,
  A.~Ziogas, A.~Koutroumpas, L.~Sakkas, K.~I. Gourgoulianis, and Z.~D. Daniil,
  \emph{Serum {V}{E}{G}{F} levels are related to the presence of pulmonary
  arterial hypertension in systemic sclerosis}, BMC Pulm. Med. (2009), 9:18.

\bibitem{pierce2001}
G.~F. Pierce, \emph{{{I}nflammation in nonhealing diabetic wounds: the
  space-time continuum does matter}}, Am. J. Pathol. \textbf{159} (2001),
  no.~2, 399--403.

\bibitem{podgrabinska2002}
S.~Podgrabinska, P.~Braun, P.~Velasco, B.~Kloos, M.~S. Pepper, D.~Jackson, and
  M.~Skobe, \emph{Molecular characterization of lymphatic endothelial cells},
  PNAS \textbf{99} (2002), no.~25, 16069--16074.

\bibitem{posnett2008}
J.~Posnett and P.~J. Franks, \emph{{{T}he burden of chronic wounds in the
  {U}{K}}}, Nurs Times \textbf{104} (2008), no.~3, 44--45.

\bibitem{reddy1995}
N.~P. Reddy and K.~Patel, \emph{{{A} mathematical model of flow through the
  terminal lymphatics}}, Med Eng Phys \textbf{17} (1995), no.~2, 134--140.

\bibitem{rofstad2014}
E.~K. Rofstad, K.~Galappathi, and B.~S. Mathiesen, \emph{{{T}umor interstitial
  fluid pressure -- {A} link between tumor hypoxia, microvascular density, and
  lymph node metastasis}}, Neoplasia \textbf{16} (2014), no.~7, 586--594.

\bibitem{roose2008}
T.~Roose and A.~C. Fowler, \emph{{{N}etwork development in biological gels:
  role in lymphatic vessel development}}, Bull. Math. Biol. \textbf{70} (2008),
  no.~6, 1772--1789.

\bibitem{rutkowski2006}
J.~M. Rutkowski, K.~C. Boardman, and M.~A. Swartz, \emph{Characterization of
  lymphangiogenesis in a model of adult skin regeneration}, Am. J. Physiol.
  Heart. Circ. Physiol. \textbf{291} (2006), H1402--H1410.

\bibitem{rutkowski2007}
J.~M. Rutkowski and M.~A. Swartz, \emph{{{A} driving force for change:
  interstitial flow as a morphoregulator}}, Trends Cell Biol. \textbf{17}
  (2007), no.~1, 44--50.

\bibitem{saaristo2006}
A.~Saaristo, T.~Tammela, A.~Farkkila, M.~Karkkainen, E.~Suominen,
  S.~Yla-Herttuala, and K.~Alitalo, \emph{Vascular endothelial growth
  factor-{C} accelerates diabetic wound healing}, Am. J. Pathol. \textbf{169}
  (2006), 1080--1087.

\bibitem{schugart2008}
R.~C. Schugart, A.~Friedman, R.~Zhao, and C.~K. Sen, \emph{{{W}ound
  angiogenesis as a function of tissue oxygen tension: a mathematical model}},
  PNAS \textbf{105} (2008), no.~7, 2628--2633.

\bibitem{scianna2013}
M.~Scianna, C.~G. Bell, and L.~Preziosi, \emph{A review of mathematical models
  for the formation of vascular networks}, J. Theor. Biol. \textbf{333} (2013),
  174--209.

\bibitem{sheikh2000}
A.~Y. Sheikh, J.~J. Gibson, M.~D. Rollins, H.~W. Hopf, Z.~Hussain, and T.~K.
  Hunt, \emph{Effect of hyperoxia on vascular endothelial growth factor levels
  in wound model}, Arch. Surg. \textbf{135} (2000), 1293--1297.

\bibitem{shi2011}
M.~Shi, J.~Zhu, R.~Wang, X.~Chen, L.~Mi, T.~Walz, and T.~A. Springer,
  \emph{Latent {TGF-$\beta$} structure and activation}, Nature \textbf{474}
  (2011), 343--351.

\bibitem{shields2007}
J.~D. Shields, M.~E. Fleury, C.~Yong, A.~A. Tomei, G.~J. Randolph, and M.~A.
  Swartz, \emph{Autologous chemotaxis as a mechanism of tumor cell homing to
  lymphatics via interstitial flow and autocrine ccr7 signaling}, Cancer Cell
  \textbf{11} (2007), no.~6, 526--538.

\bibitem{simonsen2012}
T.~G. Simonsen, J.~V. Gaustad, M.~N. Leinaas, and E.~K. Rofstad, \emph{{{H}igh
  interstitial fluid pressure is associated with tumor-line specific vascular
  abnormalities in human melanoma xenografts}}, PLoS ONE \textbf{7} (2012),
  no.~6, e40006.

\bibitem{singer1999}
A.~J. Singer and R.~A. Clark, \emph{{{C}utaneous wound healing}}, N. Engl. J.
  Med. \textbf{341} (1999), no.~10, 738--746.

\bibitem{stadelmann1998}
W.~K. Stadelmann, A.~G. Digenis, and G.~R. Tobin, \emph{{{P}hysiology and
  healing dynamics of chronic cutaneous wounds}}, Am. J. Surg. \textbf{176}
  (1998), no.~2A Suppl, 26S--38S.

\bibitem{sutton1991}
A.~B. Sutton, A.~E. Canfield, S.~L. Schor, M.~E. Grant, and A.~M. Schor,
  \emph{The response of endothelial cells to {TGF}$\beta$-1 is dependent upon
  cell shape, proliferative state and the nature of the substratum}, J. Cell
  Sci. \textbf{99} (1991), 777--787.

\bibitem{sweat2012}
R.~S. Sweat, P.~C. Stapor, and W.~L. Murfee, \emph{{{R}elationships between
  lymphangiogenesis and angiogenesis during inflammation in rat mesentery
  microvascular networks}}, Lymphat. Res. Biol. \textbf{10} (2012), no.~4,
  198--207.

\bibitem{swift2001}
M.~E. Swift, A.~L. Burns, K.~L. Gray, and L.~A. DiPietro, \emph{{{A}ge-related
  alterations in the inflammatory response to dermal injury}}, J. Invest.
  Dermatol. \textbf{117} (2001), no.~5, 1027--1035.

\bibitem{tammela2010}
T.~Tammela and K.~Alitalo, \emph{Lymphangiogenesis: Molecular mechanisms and
  future promise}, Cell \textbf{140} (2010), 460--476.

\bibitem{taylor2009}
A.~W. Taylor, \emph{Review of the activation of {TGF-$\beta$} in immunity}, J.
  Leukocyte Biol. \textbf{85} (2009), 29--33.

\bibitem{tranquillo1988}
R.~T. Tranquillo, S.~H. Zigmond, and D.~A. Lauffenburger, \emph{{{M}easurement
  of the chemotaxis coefficient for human neutrophils in the under-agarose
  migration assay}}, Cell Motil. Cytoskeleton \textbf{11} (1988), no.~1, 1--15.

\bibitem{vandenberg2003}
B.~M. van~den Berg, H.~Vink, and J.~A.~E. Spaan, \emph{The endothelial
  glycocalyx protects against myocardial edema}, Circ. Res. \textbf{92} (2003),
  592--594.

\bibitem{velazquez2004a}
J.~J.~L. Vel{\'a}zquez, \emph{{P}oint {D}ynamics in a {S}ingular {L}imit of the
  {K}eller--{S}egel {M}odel 1: {M}otion of the {C}oncentration {R}egions}, SIAM
  J. Appl. Math. \textbf{64} (2004), no.~4, 1198--1223.

\bibitem{velazquez2004b}
\bysame, \emph{{P}oint {D}ynamics in a {S}ingular {L}imit of the
  {K}eller--{S}egel {M}odel 2: {F}ormation of the {C}oncentration {R}egions},
  SIAM J. Appl. Math. \textbf{64} (2004), no.~4, 1224--1248.

\bibitem{staden1989}
H.~von Staden, \emph{Herophilus [and] the art of medicine in early alexandria},
  Cambridge University Press, Cambridge, UK, 1989.

\bibitem{vowden2011}
P.~Vowden, \emph{Hard-to-heal wounds made easy}, Wounds Intern. \textbf{2}
  (2011), no.~4.

\bibitem{wahl1987}
S.~M. Wahl, D.~A. Hunt, L.~M. Wakefield, N.~McCartney-Francis, L.~M. Wahl,
  A.~B. Roberts, and M.~B. Sporn, \emph{Transforming growth factor type $\beta$
  induces monocyte chemotaxis and growth factor production}, PNAS \textbf{84}
  (1987), 5788--5792.

\bibitem{wakefield1988}
L.~M. Wakefield, D.~M. Smith, K.~C. Flanders, and M.~B. Sporn, \emph{{{L}atent
  transforming growth factor-beta from human platelets. {A} high molecular
  weight complex containing precursor sequences}}, J. Biol. Chem. \textbf{263}
  (1988), no.~16, 7646--7654.

\bibitem{waugh2006}
H.~V. Waugh and J.~A. Sherratt, \emph{Macrophage dynamics in diabetic wound
  healing}, Bull. Math. Biol. \textbf{68} (2006), 197--207.

\bibitem{weber1990}
K.~Weber-Matthiesen and W.~Sterry, \emph{Organization of the
  monocyte/macrophage system of normal human skin}, J. Invest. Dermatol.
  \textbf{95} (1990), 83--89.

\bibitem{whitehurst2006}
B.~Whitehurst, C.~Eversgerd, M.~Flister, C.~M. Bivens, B.~Pickett, D.~C.
  Zawieja, and S.~Ran, \emph{Molecular profile and proliferative responses of
  rat lymphatic endothelial cells in culture}, Lymph. Res. Biol. \textbf{4}
  (2006), no.~3, 119--142.

\bibitem{withington1984}
E.~T.~T. Withington, \emph{Hippocrates on joints}, vol.~3, Harvard University
  Press, Cambridge, MA, 1984.

\bibitem{witte2001}
M.~H. Witte, M.~J. Bernas, C.~P. Martin, and C.~L. Witte,
  \emph{Lymphangiogenesis and lymphangiodysplasia: from molecular to clinical
  lymphology}, Microsc. Res. Tech. \textbf{55} (2001), 122--145.

\bibitem{yang2009}
J.~P. Yang, H.~J. Liu, S.~M. Cheng, Z.~L. Wang, X.~Cheng, H.~X. Yu, and X.~F.
  Liu, \emph{{{D}irect transport of {V}{E}{G}{F} from the nasal cavity to
  brain}}, Neurosci. Lett. \textbf{449} (2009), no.~2, 108--111.

\bibitem{yang1999}
L.~Yang, C.~X. Qiu, A.~Ludlow, M.~W.~J. Ferguson, and G.~Brunner, \emph{Active
  transforming growth factor-$\beta$ in wound repair -- determination using a
  new assay}, Am. J. Pathol. \textbf{154} (1999), no.~1, 105--111.

\bibitem{zachary2001}
I.~Zachary and G.~Gliki, \emph{Signaling transduction mechanisms mediating
  biological actions of the vascular endothelial growth factor family},
  Cardiovasc. Res. \textbf{49} (2001), 568--581.

\bibitem{zheng2007}
Y.~Zheng, M.~Watanabe, T.~Kuraishi, S.~Hattori, C.~Kai, and M.~Shibuya,
  \emph{{{C}himeric {V}{E}{G}{F}-{E}{N}{Z}7/{P}l{G}{F} specifically binding to
  {V}{E}{G}{F}{R}-2 accelerates skin wound healing via enhancement of
  neovascularization}}, Arterioscler. Thromb. Vasc. Biol. \textbf{27} (2007),
  no.~3, 503--511.

\bibitem{zhuang1997}
J.~C. Zhuang and G.~N. Wogan, \emph{Growth and viability of macrophages
  continuously stimulated to produce nitric oxide}, PNAS \textbf{94} (1997),
  11875--11880.

\bibitem{zimny2002}
S.~Zimny, H.~Schatz, and M.~Pfohl, \emph{Determinants and estimation of healing
  times in diabetic foot ulcers}, J. Diab. Compl. \textbf{16} (2002), no.~5,
  327--332.

\end{thebibliography}


\newpage

\tableofcontents

\end{document}